\documentclass[a4paper,11pt]{article}
\pdfoutput=1 

\usepackage{jheppub} 

\usepackage[T1]{fontenc} 
\usepackage{mathrsfs}
\usepackage{extarrows}
 \usepackage{quiver} 
\usepackage{tikz}
\usetikzlibrary{arrows.meta}
\usepackage{musicography}
\usepackage{empheq}
\usepackage{mathrsfs}
 
\usepackage{bm}
\usepackage{verbatim}
 
 \usepackage{makecell}
 \usepackage{float}
 \usepackage{hyperref}

 \usepackage[makeroom]{cancel}
 \usepackage{dsfont}
  
\newcommand{\be}{\begin{equation}}
\newcommand{\ee}{\end{equation}}
\newcommand{\bpm}{\begin{pmatrix}}
\newcommand{\epm}{\end{pmatrix}}

\newcommand{\PBK}[1]{\ensuremath{\begin{pmatrix}#1\end{pmatrix}}}

\newcommand{\EV}[1]{\langle #1 \rangle}
\newcommand{\beqn}{\begin{eqnarray}}
\newcommand{\eeqn}{\end{eqnarray}}

\newcommand{\GL}{\text{GL}}
\usepackage{slashed}

\DeclareMathOperator{\sgn}{sgn}

\DeclareMathOperator{\Tr}{Tr}
\DeclareMathOperator{\tr}{tr}

\usepackage{pgffor}
\foreach \x in {A,...,Z}{
  \expandafter\xdef\csname b\x\endcsname{\noexpand\mathbb{\x}}
}
\foreach \x in {A,...,Z}{
  \expandafter\xdef\csname c\x\endcsname{\noexpand\mathcal{\x}}
}
\foreach \x in {A,...,Z}{
  \expandafter\xdef\csname s\x\endcsname{\noexpand\mathscr{\x}}
}
\foreach \x in {A,...,Z}{
  \expandafter\xdef\csname sf\x\endcsname{\noexpand\mathsf{\x}}
}
\foreach \x in {a,...,z}{
  \expandafter\xdef\csname sf\x\endcsname{\noexpand\mathsf{\x}}
}
\foreach \x in {A,...,Z}{
  \expandafter\xdef\csname  fk\x\endcsname{\noexpand\frak{\x}}
}
\foreach \x in {a,...,z}{
  \expandafter\xdef\csname  fk\x\endcsname{\noexpand\frak{\x}}
}
\newcommand{\wt}{\widetilde}

 \newcommand{\bv}{ \begin{verbatim}}

\newcommand{\p}{\partial}
\newcommand{\sVec}{\text{sVec}}

 \newcommand{\PA}{\paragraph}

\newcommand{\diag}{\text{diag}}
 \newcommand{\SNF}{\text{SNF}}

\DeclareMathOperator{\Hom}{Hom}

\newcommand{\HB}{\text{HB}}
\newcommand{\CB}{\text{CB}}
\newcommand{\HS}{\text{HS}}

\author[\natural]{Hongliang Jiang }

 \affiliation[\natural{}]{Blackett Laboratory, Imperial College, London SW7 2AZ, U.K.}

 \emailAdd{h.jiang@imperial.ac.uk}

\title{\boldmath\Huge	  Time-reversal invariant TQFTs from \\ self-mirror symmetric SCFTs}

\abstract{
We establish a connection between three-dimensional self-mirror symmetric  $\mathcal N=4$ superconformal field theories (SCFTs) and time-reversal invariant topological quantum field theories (TQFTs) arising from universal mass deformations. Focusing on the Abelian case, the ultraviolet (UV) SCFT is characterized by the charge matrix  $Q$, while the infrared (IR) TQFT corresponds to an Abelian Chern-Simons theory with level matrix $K=QQ ^T$. We derive constraints on the charge matrix for self-mirror symmetric SCFTs and demonstrate that the Coulomb and Higgs branch Hilbert series of these theories coincide.
Additionally, we derive a general formula for the superconformal indices of Abelian  $\mathcal N=4$
  SCFTs with arbitrary charge matrices. For SCFT  with the constrained charge matrix, the superconformal index is argued to exhibit invariance under the inversion of fugacity associated with R-symmetry, providing further evidence of self-mirror symmetry.
We explore various properties of time-reversal invariant Abelian Chern-Simons theories in detail and establish their connections to self-mirror symmetry in SCFTs from multiple perspectives. In particular, we introduce a   quantity, dubbed Gauss generating function, which is real and  thus invariant under complex conjugation for time-reversal symmetric TQFTs, in parallel with  the   superconformal index, which is invariant  under  the inversion of R-symmetry fugacity for   self-mirror symmetric SCFTs. 
}

\begin{document} 
\maketitle
\flushbottom

\section{Introduction}

Superconformal field theories (SCFTs) and topological quantum field theories (TQFTs) represent remarkable classes of quantum field theories. Their additional structures and symmetries often allow for analytical study,  providing valuable insights into the broader framework of quantum field theories.  Moreover, they exhibit profound mathematical structures and are frequently interconnected with each other.   In particular, there is a   powerful procedure, called   topological twist \cite{Witten:1988ze},   which has enabled the construction of a large variety of TQFTs from  superconformal and more generally supersymmetric field  theories. One famous example is given by the Kapustin-Witten TQFT, which  is   obtained  from the 4d  $\cN=4$  super Yang-Mills theory via   topological twist \cite{Kapustin:2006pk}.  TQFT also plays an important role in characterizing the IR behavior of general quantum field theories. A general wisdom of quantum field theories is that they can be obtained from some UV conformal field theories by adding some relevant deformations.  The relevant deformation triggers the renormalization group (RG) flow, leading to various possible IR phases. While  conformal field theories (CFTs) are instrumental in analyzing  the gapless phase, TQFTs offer  the crucial tools to study the gapped phase. 
Understanding the IR phases of RG flows is a significant and intriguing question   in many  physical   systems.

In this paper, we study a special type of RG flow, which is triggered by the universal mass deformation in 3D $\cN=4$ SCFTs. Given the wide variety of $\cN=4$  SCFTs, the corresponding IR phases are anticipated to be equally diverse. 
   The  universal mass deformation gives a canonical and   universal  way to deform 3d $\cN=4$ SCFTs, because it is constructed out of stress tensor multiplet and thus exists in any local 3d $\cN=4$ SCFTs. 
\footnote{This feature closely resembles the    $T\bar T$ deformation in 2d QFT \cite{Cavaglia:2016oda,Smirnov:2016lqw}  with a key distinction:  the universal mass deformation studied in this paper is a relevant and supersymmetry-preserving deformation, while $T\bar T$ deformation is an irrelevant but integrable deformation. } Furthermore, it has been argued in \cite{Cordova:2016xhm} that the IR phase  in this case is gapped. Therefore, the   universal mass deformation gives a map from SCFTs to TQFTs in 3d. This then raises numerous interesting questions. In particular, the 3d TQFTs are fairly well-understood and generally can be described using modular tensor category (MTC) at a formal level.   They provide a very systematic  framework for understanding   anyons, which  are some quasi-particles with fractional statistics in (2+1)d.  The 3d Chern-Simons (CS) theories are important classes of  3d TQFTs.  Due to the lack of local excitations, the fundamental objects in TQFTs are the worldlines of anyons, which correspond to Wilson lines in the language of CS theory.

One distinguishing property of 3D $\cN=4$ SCFT  is mirror symmetry \cite{Intriligator:1996ex}.  The 3D $\cN=4$ SCFTs have R-symmetry 
$\fks\fko(4)\simeq \fks\fku(2) \times \fks\fku(2)$, where the two factors of $\fks\fku(2)$ are exchanged under mirror symmetry. One can obtain 3D $\cN=4$ SCFT  from UV supersymmetric gauge theory description. So it may happen that two different UV gauge theories yield two equivalent superconformal field theories up to the exchange of the two $\fks\fku(2)$ factors of R-symmetry. 

Among  the rich families of theories, we will focus on the Abelian case for simplicity. In the UV, we start with 3d $\cN=4$ Abelian SCFTs which  can be constructed from U(1) gauge vector multiplets   and  hypermultiplets with interactions governed  by the  charge matrix $Q$. It turns out that in the Abelian case, mirror symmetry boils down to a simple relation of charge matrices  for the   mirror pair.  
 In the IR, we consider Abelian TQFTs where the fusion   of anyons satisfies the multiplication rule of some Abelian group. It is believed that   all Abelian TQFTs are given by the Abelian CS theory, which can be defined through the CS action and is fully characterized by a matrix $K$. Many aspects of universal mass deformation have been studied in \cite{KNBalasubramanian:2024uvi}. In particular,  It has been argued there  that the  Abelian SCFT with charge matrix $Q$ gives rise to  the Abelian CS theory with $K$ matrix $QQ^T$  up to a sign related to the sign of the deformation parameter.  \footnote{This statement holds for SCFTs without 1-form symmetry that we will   consider  exclusively in this paper.  }
 
 Interestingly, it has also been demonstrated in \cite{KNBalasubramanian:2024uvi} that the mirror symmetry in the UV SCFTs descends to  the duality in the IR TQFTs after performing the universal mass deformation. This duality can be regarded as some generalization of the famous   level-rank duality in 
Chern-Simons theory.  The canonical example of  level-rank duality is given by $SU(N)_k \leftrightarrow U(k)_{-N}$  CS theories~\cite{Naculich:1990pa,Naculich:1990hg,Naculich:1992uf}.~\footnote{Via CS/WZW correspondence \cite{Witten:1988hf},   level-rank duality also extends to 2d CFTs described by WZW model.} \footnote{See \cite{Hsin:2016blu} for some clarifications on  level-rank duality in  CS theories as \emph{spin} theories. This also plays an important role in  the study  of dualities in  Abelian CS theories. } The relation of mirror symmetry in UV SCFT and duality in IR TQFT is illustrated in figure~\ref{mirrorscfttqft}. These  relations have been proved   using some matrix identities of charge matrix coming from mirror symmetry, together with   some rules for   testing the duality in Abelian CS theory.

The focus of this paper is to understand the special case where the UV SCFT is self-mirror symmetric. 
And the main goal   is to show that such self-mirror symmetric SCFTs give rise to time-reversal invariant TQFTs after performing universal mass deformation.  Although   the time-reversal symmetry is broken by  universal mass deformation  explicitly, it may reappear as an emergent symmetry due to its quantum nature. This indeed happens for self-mirror symmetric SCFT. This could be possible thanks to the level-rank duality as well as its generalizations. For example, $SO(N)_N$ is equivalent to $SO(N)_{-N}$ and  is thus time-reversal invariant  thanks to the level-rank duality $SO(N)_k\leftrightarrow SO(k)_{-N}$. Note that   time-reversal just flips the sign of CS level. In this sense, time-reversal invariance can be understood as a special case of the generalized level-rank duality. In the   self-mirror symmetric case, the mirror pair of two SCFTs, which is essentially the SCFT and itself, yield two IR CS theories with opposite $K$ matrices.  Further using the previously established results on the duality of the  two IR CS theories arising from UV mirror symmetry, we learn  that  the two IR CS theory with opposite $K$ matrix are actually dual to each other.  Since flipping the sign of level is just the action  of time-reversal, we finally conclude that   self-mirror symmetric SCFTs yield   time-reversal invariant TQFTs.   

Building on this intuition, we will discuss  self-mirror symmetric SCFTs and time-reversal invariant TQFTs in detail, and establish their connections from several different perspectives: 

\begin{itemize}
\item Firstly, we will give a systematic construction of Abelian self-mirror symmetric SCFTs. This turns out to impose some constraints on the charge matrix, which arise  from the matrix identity from mirror symmetry and the redundant description of the SCFT using charge matrix.
To further justify the constraints, we study the Hilbert series and superconformal indices for these theories. We will show that those constraints on charge matrices actually guarantee that the Coulomb and Higgs branch Hilbert series are the same. We will further derive a general formula of   superconformal index for SCFT with arbitrary charge matrix, whose special limits reproduce the  Coulomb branch (CB) and Higgs branch (HB) Hilbert series. Again we will show the  superconformal indices are invariant under the inversion of R-symmetry fugacity once accounting for the constraint on the charge matrix. 

 Although we will not try to find the most general solution to the constraints on  the charge matrix, we will illustrate several  infinite series of solutions, corresponding to several families of self-mirror symmetric SCFTs. In particular, in the rank-1 case with a single U(1) gauge node, we will construct  all self-mirror symmetric SCFTs, which turn out to be classified by two coprime integers. 
 
 \item
 
  Secondly, we will discuss many aspects of   time-reversal invariant abelian CS theory. Many results on this question have been established in  \cite{Geiko:2022qjy}. \footnote{See also \cite{Lee:2018eqa,Delmastro:2019vnj}   for  other discussions.} We will review them from a physical  perspective, clarify their connections, and illustrate with many explicit examples. In particular, we will show several different methods to  determine whether the Abelian CS theory is  time-reversal invariant.  The first method arises as a special case of  duality as we discussed above; the second method is related to the self-perpendicularity   of some  lattice underlying the classical CS data; the third method boils down to the classification of $\sfT$-symmetric quadratic form underlying the quantum CS data; the fourth method is computationally straightforward and related to the reality of Gauss sums or their corresponding generating functions, which will be referred to as Gauss generating function.
  
   For U(1) CS theory with arbitrary level, it turns out that the time-reversal invariance holds if and only if the level  can be expressed as  the sum of squares of two coprime integers.

\item Finally, we will relate  the  self-mirror symmetry  in UV and time-reversal symmetry in IR.  More specifically, we will prove  that   self-mirror symmetric SCFTs give rise to  time-reversal symmetric TQFTs in the IR after performing the universal mass deformation. Several different methods will be provided to establish the connection.  In particular, we will give some explicit examples by computing superconformal indices and Gauss generating functions, which  are the most straightforward   ways to test the self-mirror symmetry and time-reversal invariance.  \footnote{This computational approach may   be   useful in the future   generalization   to   non-Abelian case.   Note that despite some positive evidences  \cite{Geiko:2022qjy},  it remains  a conjecture that the non-Abelian   CS theory is time-reversal invariant if and only if the Gauss sums are all real. }

\end{itemize}

The rest of the paper is organized as follows.
In section  \ref{universalmassdef}, we review some general properties of universal mass deformation. We also introduce some basic properties of TQFTs and SCFTs that will be studied later on. We further show that  mirror symmetry in SCFT descends to the generalized level-rank duality in TQFT after performing the universal mass deformation.
In section \ref{SMSCFT}, we  will discuss    how to construct self-mirror symmetric Abelian SCFTs, and show that  their   Hilbert series and superconformal indices satisfy the   transformation properties expected from self-mirror symmetry. 
In section \ref{TinvTQFT}, we will discuss extensively the time-reversal invariance of Abelian CS theory. 
The time-reversal invariance can be discriminated    equivalently from several  methods, including the reality of Gauss sums/generating function, the self-perpendicularity of lattices, and satisfaction of matrix identities for duality. We will give some simple examples to illustrate these different methods. 
In section \ref{SCFTtoTQFT}, we  establish   the main claim of this paper, namely self-mirror symmetric   SCFTs give rise to time-reversal invariant TQFTs.
In section \ref{conclusion}, we summarize the main results and discuss  some open questions for future explorations.

      \section{Universal mass deformation in 3d $\cN=4$ SCFT}\label{universalmassdef}
      In this section, we will review some general properties of universal mass deformation in 3d $\cN=4$ SCFTs.  Following\cite{KNBalasubramanian:2024uvi}, we will  then discuss the IR phases, given by certain  TQFTs. In particular, we will show that the mirror pair of two SCFTs give rise to two TQFTs in the IR which are dual to each other.  For simplicity, we will focus on the Abelian case for both SCFTs and TQFTs.
      
    \subsection{General aspect  of universal mass deformation}   \label{scftnotation}
      The  superconformal algebra underlying any 3d $\cN=4$ SCFT  is given by $\fko\fks\fkp(4|4)$, whose bosonic sub-algebra is $\fks\fko(4,1) _\text{conf}\times \fks\fko(4)_R $.  
        The former accounts for the conformal symmetry  including Lorentz symmetry $\fks\fko(3)_L\simeq \fks\fku(2)$ and dilation symmetry $\fks\fko(1,1)_D\simeq \fkr$  (in addition to special conformal symmetry), while the latter corresponds to the   $R$-symmetry    $\fks\fko(4)_R \simeq \fks\fku(2)\times  {\fks\fku(2)}'$. As a result, we can represent each conformal multiplet  as $[j]_\Delta^{(R,R')}$, where $\Delta$ is the scaling dimension, $j, R, R'\in \bZ_{\ge 0}$ are    the Dynkin labels for various $ {\fks\fku(2)}$  representations     of Lorentz symmetry and R-symmetries. All local operators in the   multiplet can then be written as $\cO_ {(\alpha_1 \cdots  \alpha_{  j })(a_1 \cdots  a_{ R })(\dot a_1 \cdots  \dot a_{ R' })}$, where $\alpha_i=\pm,a_i, \dot a_i=1,2$. 
       
 In SCFTs,   the local operators are further organized in  superconformal multiplets. In particular, the stress-tensor multiplet  universally appears in all local SCFTs.  For 3d $\cN=4$ SCFTs, the  stress-tensor multiplet has the following structure \cite{Cordova:2016emh}
        \be\label{STmul}
    A_2[0]_1^{(0,0)} \xlongrightarrow{Q} [1]_{\frac32}^{(1,1)}
\xlongrightarrow{Q} [2]_2^{(0,2)}\oplus [2]_2^{(2,0)} \oplus [0]_2^{(0,0)}
    \xlongrightarrow{Q} [3]_{\frac52}^{(1,1)}
    \xlongrightarrow{Q} [4]_3^{(0,0)}~,
    \ee
where each entry in the sequence denotes the conformal primaries related by the actions of Poincare supercharges, and the first entry above  is the superconformal primary, annihilated by all   superconformal supercharges.  Here $A_2$ denotes a specific  type of shortening condition  associated to the stress-tensor multiplet \cite{Cordova:2016emh}.
 
 If we denote the superconformal primary of the stress-tensor multiplet as $\cJ$, then the shortening condition is given by  
\be 
Q_{\alpha a \dot a}Q^{\alpha}{}_{  b \dot b}\cJ=\frac14\epsilon_{ab}\epsilon_{\dot a\dot b}Q_{\alpha c \dot c}Q^{\alpha c \dot c}\cJ,\qquad
   \cJ\in A_2[0]^{(0;0)}_1 
\ee
where $Q_{\alpha a \dot a} $ are the Poincare supercharges.

We can further   look     at the scalar primary operator in the level-2 super-descendant   \eqref{STmul}
\be
J=Q_{\alpha a \dot a}Q^{\alpha a \dot a}\cJ \quad \in\;  [0]_2^{(0,0)}~,
\ee
If we   act on it with  Poincare supercharges,  one can argue that it gives a total derivative: 
\be\label{QJ}
Q_{\gamma c\dot c}J=Q_{\gamma c\dot c}Q_{\alpha a \dot a}Q^{\alpha a \dot a}\cJ  \sim \p X~,
\ee
This can be easily argued by contradiction: if we don't get a total derivative, then we should get a conformal   primary operator which appears as the  level-3 descendant in \eqref{STmul}. The unique level-3 conformal primary there is given by $[3]_{\frac52}^{(1,1)}$, diagreeing with $[1]_{\frac52}^{(1,1)} $ that one would expect from acting   Poincare supercharge on $J$ which is a singlet under R-symmetry  and Lorentz symmetry.

Now we can use the operator $J$ to deform the SCFT 
\be\label{unidef}
\delta S=  \int d^3 x\; \delta \cL=m \int d^3 x\;   J+\cdots= m \int d^3 x\;   Q_{\alpha a \dot a}Q^{\alpha a \dot a}\cJ+\cdots~,
\ee
where the ellipsis stands for terms at higher non-liner order. 
Thanks to the property  \eqref{QJ},   this deformation preserves supersymmetry   
\be
Q_{\gamma c\dot c} \delta S \sim m \int d^3 x\;  \p X =0~.
\ee
  We will refer to  the deformation \eqref{unidef} by $J$ as universal mass deformation, because it  is built from the universally existing stress-tensor multiplet. Since $J$ has scaling dimension 2, the universal mass deformation is a relevant deformation, which can change the IR phase under the RG flow. 

 The universal mass deformation has several interesting effects. 
  In particular,  it modifies the supersymmetry algebra  \cite{Cordova:2016xhm}
\be\label{QQalgebra}
\{Q_{\alpha a \dot a}, Q_{\beta  b \dot b}\}=\epsilon_{ab}\epsilon_{\dot a\dot b}P_{(\alpha\beta)}
+ m \epsilon_{\alpha \beta} \Big( \epsilon_{ab}R'_{\dot a\dot b}
-  \epsilon _{\dot a\dot b} R _{ab} \Big)~,
\ee
 where the R-symmetry generators on the RHS give rise to the non-central extension of the standard supersymmetry  algebra.  The non-central extension is a very special feature in 3D (and also 2D) and it is not allowed in higher dimensions. This kind of  unconventional supersymmetry algebra     was presented  explicitly previously in  \cite{Lin:2005nh}. Starting from the non-central extended algebra \eqref{QQalgebra}, it has been argued in \cite{Lin:2005nh,Cordova:2016xhm} that  all  the massless representation  must be  trivial with zero energy,  hence the only massless single-particle state is just  given by the vacuum.  Therefore, the IR theory  must be gapped and described by some possibly trivial  TQFT.
  For example, adding the universal mass deformation to the free hypermultiplet triggers   mass terms for all the fields, which leads to a gapped theory with massive scalars and fermions. 
  
 A prominent feature  of 3d $\cN=4$ SCFTs is mirror symmetry under which the $\fks\fku(2)$ factors  of $\fks\fko(4)_R$ symmetry are exchanged.   The mirror symmetry corresponds to an automorphism of   the algebra \eqref{QQalgebra}  acted by $R\leftrightarrow R', m\leftrightarrow -m$. 
 
     \subsection{3d $\cN=4$ Abelian SCFT and mirror symmetry}  
     \subsubsection{Charge matrix}\label{chargematrx}
     We will focus on a special type of SCFTs which arise as the IR fixed points of some 3d $\cN=4$ supersymmetric gauge theories. The supersymmetric gauge theory consists of the $N$ hypermultiplets and $k$ vector multiplets subject to $k\le N$. We also need to specify the charges of hypermultiplets under each vector multiplet. This   can be described in terms of  gauge charge matrix    $Q_i^a$, where $i=1, \cdots, N$ labels the hypermultiplet and $a=1, \cdots,  k$ labels the vector multiplet. In addition to the topological symmetry $U(1)^k$ from each abelian vector multiplet, the resulting theory also has $U(1)^{N-k}$ flavor symmetry with flavor charge matrix $q_i^b$,  where $i=1, \cdots, N, \; b=1, \cdots, N-k$. 
     We can combine the gauge and flavor charge  matrix together and get an $N\times N$ square matrix \footnote{All the matrices in this paper have integer entries, except for those of the form   $K^{-1}$ which is the inverse of an integral matrix.}
     \be
{     \bf Q}=\PBK{ Q \\ q}, \qquad\qquad Q\in \bZ^{k\times N}~, \qquad q\in \bZ^{(N-k)\times N}~.
     \ee
We will further   require that   the matrix $\bf Q$    is unimodular, denoted as $|\det \bf Q|=1$ or ${\bf Q }\in \GL(N, \bZ)$. Physically, this means that the corresponding  3d SCFT  has no 1-form symmetry. 

Note that the charge matrix depends on the choice of basis for gauge and flavor symmetries; after performing a change of basis, the charge matrix transforms as 
\be\label{Qtsf}
{ \bf Q}=\PBK{ Q \\ q}\mapsto \PBK{ A & 0 \\ C & D}\PBK{ Q \\ q}=\PBK{ AQ \\ CQ+Dq}, \qquad 
 A\in \GL(k,\bZ), \quad D\in \GL(N-k, \bZ), \quad C \in \bZ^{(N-k)\times k}~,
\ee    
  where   $A, D$    are unimodular square matrices of size $k$ and $N-k$, respectively, and $C$ is an arbitrary integer matrix.  In particular, we see the unimodularity of $\bf Q$ is maintained under the change of basis.  
  
  Furthermore, we now show that  one can  determine the flalvor charge matrix $q$ from gauge charge matrix $Q$, up to a change of basis.  Namely, we can complete the matrix $Q$ to a unimodular square matrix $\bf Q$. 

  To see this, let us   consider the Hermite  decomposition of $Q$ \cite{DBLP:journals/ajc/MagliverasTW08}:
    \be\label{QVH}
    QV=H~,
    \ee
   where $Q\in \bZ^{k\times N } , V \in \bZ^{N\times N },H\in \bZ^{k\times N } $. Here $V$ is unimodular $|\det V|=1$, while  $H$ is called the Hermite Normal Form and  takes the form 
      \be
   H=\PBK {G  & \quad 0_{k \times (N-k)}}~, \qquad G\in \bZ^{k\times k } ~,
   \ee
   where the square matrix $G $ is lower triangular  with the property that its main diagonal entries are positive and the value of any of its off-diagonal entries below the main diagonal is between 0 and the value of the diagonal entry in the same row.       Note the matrix $H$ is unique.
   
    Let $q$ denote the matrix formed by the last  $N-k$ rows of $V^{-1}$.  Then one can combine $Q$ and $q$ to form a square matrix $\bf Q=\PBK{ Q \\q }$ and furthermore  show that $|{\det \bf Q}|=|\det G|$ \cite{DBLP:journals/ajc/MagliverasTW08}.     If all the diagonal entries of $G$ is 1, then $\bf Q$  is a unimodular square matrix, meaning that  $Q$ can be extended to a unimodular matrix $\bf Q$. In this case, we  refer to  the matrix $Q$ as a primitive matrix.  This plays an important role. 
    
      There are several equivalent  ways to test    the primitivity of    $Q$. More precisely, the matrix $Q$ is   primitive  if one of the     following equivalent conditions is  satisfied:
    \begin{itemize} 
  \item[1.] All  the diagonal entries of the    Hermit Normal Form $H$  of $Q$ in  \eqref{QVH}  are  1.
  \item[2.] The matrix $Q$  can be extended or completed to a unimodular matrix.
    \item[3.] The  greatest common divisor of all   maximal minors of $Q$  is  1 \cite{newman1972integral}.     More specifically, we can take all possible $k$ columns of $Q$ to  form square matrices, and compute their determinants $d_i$ where $i=1,\cdots,   \frac{ n! }{ k! (n-k)!}$. Then the condition is   $ \gcd (d_1, \cdots )=1$.  
    
     \end{itemize}     
    
In this paper, we will only study the case that $Q$ is primitive. If necessary, we can easily complete the primitive $Q$ to a  unimodular matrix $\bf Q$ using the Hermitian normal form of $Q$. We will use $ T_Q$ to denote the UV supersymmetric gauge theory constructed from   $k$ abelian vector multiplets and $N$ hypermultiplets with couplings specified by primitive charge  matrix $Q\in \bZ^{k\times N}$, and the corresponding SCFT at the IR fixed point will be denoted as   $\cT_Q$. 
  
     \subsubsection{Mirror symmetry} 
     Now we would like to understand the mirror symmetry in Abelian SCFTs.  As we discussed before, the SCFT $\cT_Q$ has topological symmetry   $U(1)^k$ and flavor symmetry $U(1)^{N-k}$ from Coulomb and Higgs fields, respectively. Since mirror symmetry exchange the Coulomb and Higgs, we expect the mirror dual of $\cT_Q$ should have topological symmetry   $U(1)^{N-k}$ and flavor symmetry $U(1)^{ k}$, implying that it is given by $\wt k\equiv N-k$ Abelian vector multiplets and $N$ hypermultiplets. The charge matrix of the mirror  theory  is simply given by 
     \be\label{QQtilde}
     \widetilde    {  \bf Q}=\PBK{ \widetilde q \\ \widetilde Q}=({ \bf Q}^T)^{-1}~,
      \qquad\qquad\qquad 
      \wt Q\in \bZ^{(N-k)\times N}~, \qquad q\in \bZ^{k\times N}~,
     \ee
  where $\widetilde Q_i^b$  with  $i=1, \cdots, N,\; b= 1, \cdots N-k$,  and $\widetilde q_i^a$  with  $i=1, \cdots, N, \; a= 1, \cdots  k$.  They are gauge and flavor charges, respectively. 
  Obviously, if $\bf Q$ is unimodular, then  $\bf \wt Q$ is also unimodular,   $ |\det \widetilde    {  \bf Q}|=1$.
  Therefore mirror symmetry preserves the unimodularity  or primitivity of charge matrix. 
  
  Let us look into   the  charge matrices in more detail  and prove some useful identities.  From \eqref{QQtilde}, we learn that 
  \footnote{We will use $\mathds{1}_n$ to denote the identity matrix of size $n \times n$.}
   \be 
      {   \bf Q}^T   {   \widetilde{  \bf Q}}= Q^T\wt q + q^T\wt Q    =\mathds{1}_N~,
    \qquad
            {    {   \widetilde{  \bf Q}  \bf Q}^T  }
            = \PBK{\wt q Q^T &\wt q q^T\\ \wt Q Q^T &\wt Q q^T } =\mathds{1}_N~,
  \ee 
    which in particular implies that 
    \be\label{mirrorId}
    Q^T\wt q + q^T\wt Q    =\mathds{1}_N~, \qquad
    \wt q Q^T  = \mathds{1}_{ k}~, \qquad
    \wt Q q^T  =\mathds{1}_{N-k}~, \qquad
\wt q q^T=0, \qquad \wt Q Q^T=0~.
    \ee 
   We can also take the transpose of the above identities, and get a new set of identities, which can be equivalently obtained from the exchange $Q\leftrightarrow \wt Q$ and $q \leftrightarrow \wt q$: 
     \be\label{mirrorId2}
   \wt  Q^T  q +\wt  q^T  Q    =\mathds{1}_N~, \qquad
      q \wt Q^T  = \mathds{1}_{N- k}~, \qquad
      Q \wt q^T  =\mathds{1}_{ k}~, \qquad
  q \wt q^T=0, \qquad   Q \wt Q^T=0~.
    \ee

   As we discussed, the flavor charge matrices can be obtained from the gauge charge matrices. Therefore, the most important identity for mirror symmetry is supposed to be given by the last equality in \eqref{mirrorId}: $ \wt Q Q^T=Q\wt Q^T=0$. If we regard each row of the charge matrix $Q$ as the coordinates of vectors, namely 
   \be
   Q=\PBK{v_1 \\ v_2 \\ \vdots \\ v_k }, \qquad \wt Q=\PBK{\wt v_1 \\ \wt v_2 \\ \vdots \\  \wt v_k },
   \ee
  then all the vectors $v_i$   span a  $k$-dimensional  subspace $W$ of $\bZ^N$,  and  $\wt v_i$   span a   $(N-k)$-dimensional  subspace $\wt W$ of $\bZ^N$. It is easy to see that   $v_i \cdot \wt v_j =0$. Therefore $v\cdot \wt v=0$ for any $v\in W, \wt v\in \wt W$. This means that the two vector spaces are orthogonal $W\perp \wt W$. This is called Gale duality. 
  
        \subsection{3d Abelian TQFT   and duality}
        We  now switch the gear and review some properties TQFTs, or more precisely Abelian CS theories. In particular, we would like to understand the condition under which   two Abelian CS theories  are dual to each other.

  \subsubsection{Abelian CS theory} \label{abeCStheory}
  A TQFT in 3d can be regarded as a finite collection of anyons, which are particles with fractional statistics, endowed with some additional data. The set of anyons  is denoted as $\cA$, and there is a commutative, associated product  $\times:\cA\times \cA \to\cA$ describing the fusion of anyons: 
  \be
  a \times b =\sum_c N_{ab}^c c~, \qquad a,b,c\in \cA~,
  \ee
 where the fusion coefficients $N_{ab}^c  $    are restricted to be non-negative integers in general TQFTs.
 The number of anyons in TQFT is called the rank, and is denoted as $|\cA|$.
  In this paper, we will only study the Abelian TQFT where the coefficients  $N_{ab}^c  $  can only be 0 or 1 and $\sum_c N_{ab}^c  =1$. More precisely, there is only one non-zero term in the fusion product above: $a\times b =c$ where $c=c(a,b)$ is uniquely determined. 
  In this case the set of anyons $\cA$ forms a finite Abelian group, and the fusion prdocut $\times$ is just the group multiplication on $\cA$. An Abelian TQFT is then fully specified by the finite Abelian group $\cA$ encoding the fusion of anyons, and the topological spin $\theta:\cA \to U(1)$. The Abelian group is also the one-form symmetry group of the TQFT.  All additional data of TQFT can then be   determined. The braiding between two anyons is given by
 \be
 B(a,b) =
 \frac{\theta(a\times b)}{\theta(a) \theta(b)}~,
 \ee
 while the $S$ and $T$ matrices are given by 
 \be
 S_{ab}=\frac{B(a,b)}{\sqrt{|\cA|} }~, \qquad  T_{ab}=\theta(a) e^{-\frac{2\pi i \sigma}{24}}\delta_{ab}~,
 \ee
 where $\sigma$ is the chiral central charge, which controls the framing anomaly.  Therefore the data   $(\cA, \theta,\sigma)$  completely determines the abelian  TQFT. \footnote{
 Note that for bosonic TQFT, $\sigma \mod 8$ is actually determined by Gauss sum \eqref{MGausssum}. Physically, this can be understood from the fact one can stack the TQFT with Kitaev's $E_8$ model which has central charge 8 but only  trivial anyons.  }   If the $S$ matrix is non-degenerate, the TQFT is said to be bosonic or non-spin, which can be defined on arbitrary three-manifold without requiring to specify a spin-structure. In this case, the topological spin $\theta $ is a quadratic, homogeneous function on $\cA$.

 In contrast, a spin TQFT is only well-defined on   three-manifold with a choice spin structure.    In spin TQFT, there is distinguished anyon $\psi$ with topoloigcal spin $\theta(\psi)=-1$ and trivial braiding all other anyons, namely $B(a,\psi) =1$ for all $a\in \cA$.  This implies that $\psi\times \psi=\bf 1$ where $\bf 1$ is the trivial line, and $\theta(a \times \psi) =-\theta(a)$ for all $a\in \cA$. As such, the anyon $\psi$ is called transparent fermion.  In particular, the $S$ matrix is   degenerate  for spin TQFT. The spin structure dictates the coupling of transparent fermion with three-manifold.
   
 The transparent fermion plays an important role in duality, as many dualities are only true at the level of spin TQFT \cite{Hsin:2016blu}. 
  An invertible spin TQFT has anyons $\{1,\psi \}$ where $\psi$   is the transparent fermion line with $\theta(\psi)=-1 $. It can be described by CS theory $SO(L)_1$ with central charge   $c=L/2$, where $L\in \bZ$ is defined mod 16. It is equivalent to $-L\, \text{CS}_\text{grav}$ via level-rank duality; the transparent fermion $\psi$   is then  identified with a gravitational line. Under the stacking $SO(L)_1 \times SO(  L')_1 \leftrightarrow SO(L+L')_1 \leftrightarrow -(L+L') \text{CS}_\text{grav}$. Note that  $SO(1)_1$ is the Ising  spin theory,  and $SO(2)_1 =U(1)_1$.  
    The underlying category is denoted as \sVec, which admits the 16 realizations described above.~\footnote{Note that $\sVec$  is not   modular, but admits 16-fold minimal modular extensions, which can be represented  as $Spin(L)_1$ non-spin CS theories, and   regarded as the bosonization of $SO(L)_1$ spin CS theories.   See \cite{Seiberg:2016rsg} for more details.  }   Importantly, tensoring a bosonic TQFT and $\sVec$ or equivalently $\{1,\psi \}$     yields a spin TQFT.  In the following, we will use $\{1, \psi \}$  and $\sVec$ interchangablly.  And when we refer to  a bosonic TQFT as a spin theory,  the actual spin TQFT should be understood as    the  tensoring  of the bosonic TQFT with $\sVec$.
    

 It is generally believed  that every abelian TQFT, both bosonic and spin,  in 3d can be described in terms of Abelian CS theory. The classical action of the general Abelian CS theory with gauge group $U(1)^n$ is given by \footnote{As usual, the gauge potential is normalized such that $   \frac{1}{2\pi} \oint F_i\in \bZ$ where $F=dA$ is the field strength.}
\be
S=\frac{1}{4\pi}\int_{M_3} K_{ij}A_i dA_j ~,
\ee
which is fully specified by an $n\times n $ integral   symmetric  matrix $K=K^T\in \bZ^{n \times n}$. The corresponding CS theory will be denoted as  $\mathscr T_K$. It is a bosonic TQFT if  all the diagonal entries of $K$ are even, otherwise  it is a spin TQFT and depends on the spin structure of three-manifold.
  The central charge is given by $\sigma=\sigma_+-\sigma_-$ where $\sigma_\pm$ are the number of positive and negative eigenvalues of $K$.
 
  The observables in CS theory are given by Wilson loops which can be regarded as the worldlines of anyons in TQFT
  \be
  W_{\bm \alpha}(\gamma)=\exp\Big(  i\int _\gamma \sum_i \alpha_i A_i \Big)~, \qquad \bm\alpha \in \bZ^n~.
  \ee
   
The Wilson loops are subject to some equivalence conditions.  More precisely,  the anyons in TQFT are given by the equivalence class 
\be
[\bm\alpha]\in \bZ^n /\sim~,
\ee
 where the equivalent relation  is defined such that  $\bm\alpha\sim \bm\beta$ if and only if  \cite{Delmastro:2019vnj}
  \be\label{equivclass}
\bm \alpha-\bm\beta=K\bm\gamma~, \qquad\qquad   \bm \gamma\in \bZ^n ~,\qquad \sum_ iK_{ii}\gamma_i\in 2 \bZ~.
 \ee
 So the diagonal entries of $K$ play an important role:
 \begin{itemize}
 \item  If all the diagonal entries of $K$ are even, then the TQFT is bosonic. In this case, the last condition in  \eqref{equivclass} is trivial, and the anyons are just labelled by $[\bm\alpha]\in \bZ^n /K \bZ^n$.  The total number of anyons, called rank, is $|\cA|=|\det K|$. And $\cA\cong\SNF(K) $ which is the Smith Normal Form of $K$.
  \item  If $K$  has  at least one odd  diagonal entry, then  the TQFT is spin. It  contains a transparent  fermion, and depends on the choice of spin structure. In this case, the last condition in  \eqref{equivclass} is non-trivial. The rank  is given by $|\cA|=2|\det K|$, where the factor 2 comes from the transparent fermion. 
  And $\cA\cong\SNF(K)\times \bZ_2$ where  $\bZ_2$ is generated by transparent fermion.
  
 \end{itemize}
 
 The  topological  spin in $\mathscr T_K$ CS theory is given by
    \be\label{topspins}
    \theta(\bm \alpha )=\exp(2\pi i h_{\bm \alpha})=\exp\Big( \pi i  \bm \alpha^T K^{-1}\bm \alpha \Big) ~, \qquad \bm \alpha \in \bZ^n~,
        \ee
which also determines  the   braiding  phase
    \be
     B(\bm \alpha, \bm \beta)=\frac{  \theta(\bm \alpha+\bm \beta )}{  \theta(\bm \alpha )  \theta(\bm \beta )}
     =\exp\Big(2\pi i  \bm \alpha^T K^{-1}\bm \beta \Big) ~,\qquad \bm \alpha, \bm \beta\in \bZ^n~,
    \ee

  One can easily show that the topological spin only depends on the equivalence class, namely 
      $
    \theta(\bm \alpha+K\bm\gamma ) =    \theta(\bm \alpha  )  
$
for $\bf\gamma$ satisfying $  \sum_ iK_{ii}\gamma_i\in 2 \bZ$ in \eqref{equivclass}.

   \subsubsection{Duality  in  CS theory} \label{CSduality}
        \PA{Integral lift of identity map in one CS theory.}  
Given any abelian TQFT, there is always an identity map  $\sf{id}: \cA \to\cA$ such that $[\bm\alpha]\mapsto [\bm\alpha]$. This is a  trivial map for anyons, but it may become  non-trivial if we lift it to  a linear map between integral vectors $\bm\alpha$.  At the level of integral vectors, the condition is
\be
[\bm\alpha]+ K\bm\gamma \mapsto  [\bm\alpha]+ K\bm\gamma'~, \qquad
 \sum_ iK_{ii}\gamma_i \in 2\bZ ~, \qquad  \sum_ iK_{ii}\gamma_i'\in 2\bZ ~,
\ee
for $[\bm \alpha] \in \bZ^n/\sim $ and $ \bm \gamma, \bm \gamma' \in \bZ^n$. This is  equivalent to
\be
\bm\alpha \mapsto  \bm\alpha+ K\bm\gamma ~, \qquad\quad
 \sum_ iK_{ii}\gamma_i\in 2\bZ ~,
\ee
for all $\bm \alpha \in \bZ^n$. Since the map is linear,    it should be implemented by a matrix $X$ 
such that $\bm\gamma =X\bm\alpha$ and
\be
 \sum_i K_{ii}  \gamma_i= \sum_i K_{ii}    ( X\alpha)_i= \sum_{i,j } K_{ii}   X_{ij} \alpha_j\in 2\bZ~,
\ee
 for all $\alpha_j \in \bZ$.  This means 
\be
 \sum_{i } K_{ii}   X_{ij} \in 2\bZ~.
\ee

 As a result,   the integral lift of identity map $\bm \alpha \mapsto F\bm\alpha$ is realized through 
\be\label{Fmap}
F=\mathds 1+KX~, \qquad  \sum_{i } K_{ii}   X_{ij} \in 2\bZ~, \quad \forall j~.
\ee
     
   \PA{Duality between two CS theories.}      Now we can try to understand the duality between  two Abelian CS theories described by $ {K_1}, {K_2}$ matrices.  
            A homomorphism from $\mathscr T_{K_1}$ to $\mathscr T_{K_2}$ is a linear map $\sfU: \cA_1 \to \cA_2$ such that  $\sfU$ preserves the fusion rules and topological spins
            \beqn \label{fusionpre}
            \sfU(a\times_{\cA_1} b) &=& \sfU(a) \times_{\cA_2} \sfU(b)  ~,\qquad \forall a,b \in \cA_1~,
         \\  \label{toppre}
            \theta_{\cA_1}(a) &=& \theta_{\cA_2}(\sfU(a)) ~, \qquad \forall a \in \cA_1~.
            \eeqn

  We say two TQFTs  $\mathscr T_{K_1}$ and $ \mathscr T_{K_2}$ are dual to each other if there is an isomorphism  between  $\mathscr T_{K_1}$ and $ \mathscr T_{K_2}$, namely   homomorphisms $\sfU:  \cA_1 \to \cA_2$  and $\sfU':  \cA_2 \to \cA_1$, such that  their compositions $\sfU\circ \sfU'=\sf{id}_2$ and $\sfU'\circ \sfU=\sf{id}_1$  are identity maps.

              As before, we need to lift the map $\sfU$ acting on anyons to a map acting on the integral lattice
                    \be
            \sfU: [\bm\alpha_1] \mapsto [\bm\alpha_2]~, \qquad
    [     \bm    \alpha_1]+K_1 \bm\gamma_1\mapsto[\bm \alpha_2]+K_2\bm\gamma_2~.
              \ee  
Since the map is linear, it should be implemented by a $|\cA_2|\times |\cA_1| $ matrix $R'$
             \be
                [     \bm    \alpha_2]+K_2 \bm\gamma_2= R' (   [  \bm    \alpha_1]+K_1\bm\gamma_1)~.
             \ee
 Similarly the map $\sfU'$ should be lifted to a  $|\cA_1|\times |\cA_2| $  matrix $R $. 
 
 Since we require that the compositions of $\sfU,\sfU'$ are identity maps, their integral lifts should satisfy  
           \be
          R'R =1+K_2X'~, \qquad R  R'=1+K_1X~,
          \ee
where $X$ and $X'$ should satify
     \be\label{FM2}
\sum_{i} K_{1\;ii}   X_{ij}\in 2 \bZ~, \qquad  \qquad
\sum_{i} K_{2\;ii}   X'_{ij}\in 2 \bZ ~. \qquad 
\ee
The two conditions above follow  from \eqref{Fmap}, and guarantee that the fusion rules are preserved \eqref{fusionpre}.
 
To further preserve the topological spins  \eqref{toppre}, we need to impose
   \be\label{dualityidentity}
  R^T K_1^{-1} R - K_{2}^{-1}= P~,\qquad P_{ii}\in 2\bZ~,\quad \forall i~,
      \ee
and 
   \be\label{dualityidentity}
  R'^T K_2^{-1} R' - K_{1}^{-1}= P'~, \qquad P'_{ii}\in 2\bZ~,\quad \forall i~,
      \ee
  Using \eqref{topspins},   one can verify explicitly that the topological spins are indeed invariant under the map $R, R'$.

    Collecting all equations above together, we conclude that $\mathscr T_{K_1}$ and $ \mathscr T_{K_2}$ are dual to each other if and only if  there exist  integral matrices $R,R',P, P',X,X'$ satisfying \footnote{Rigorously speaking, for two theories to be dual to each other, they must also have the same chiral central charges. However, in this paper,   we ignore   such a kind of condition, as we can easily remedy the mismatch  of chiral central charge by stacking with some invertible field theories. 
    }
%
%
%
%
         \begin{subequations}\label{dualityconditions}
         \begin{align} 
     \sfD1: &&      R  R'=1+K_1X~, \qquad \sum_{i} K_{1\;ii}   X_{ij}\in 2 \bZ~, \\
     \sfD2: &&       R'R =1+K_2X'  ~,\qquad \sum_{i} K_{2\;ii}   X'_{ij}\in 2 \bZ~,  \\
    \sfD3: &&          R^T K_1^{-1} R - K_{2}^{-1}= P~, \qquad P_{ii}\in 2 \bZ ~,\label{dualityconditions3} \\
       \sfD4: &&            R'^TK_{2}^{-1}R' -K_{1}^{-1}=P'~,  \qquad P'_{ii}\in 2 \bZ~. 
                \end{align}
         \end{subequations}
          
 Note however that the four conditions above are not all independent. For example, one can prove that  the condition $\sfD$4 above can be derived from the rest of three conditions: starting from $\sfD 3$ and using $\sfD 1$, we can   consider
         \beqn
     R'^T PR'       &  =& R'^T  R^T K_1^{-1} RR' - R'^TK_{2}^{-1}R' 
   \\   &  =&(1+K_1X)^T  K_1^{-1}(1+K_1X)- R'^TK_{2}^{-1}R'
        \\     &  =&K_1^{-1}  + X^TK_1X+X+ X^T - R'^TK_{2}^{-1}R'~,
         \eeqn
         which indeed coincides with $\sfD 4$ with $P'=  X^TK_1X+X+ X^T-     R'^T PR'   $. The diagonal entries are then given by 
                 \be\label{evencds}
      P'_{ii} \stackrel{\mod 2\;\;\;\,}{=\!=\!=\!=} \sum_j R'_{ji}R'_{ji}P _{jj} +\sum_{j } X_{ji}X_{ji} K_{1jj} \stackrel{\mod 2\;\;\;\,}{=\!=\!=\!=}  \sum_j R'_{ji} P _{jj} +\sum_{j } X_{ji}  K_{1jj} ~,
      \ee   
      which is indeed even if we use the evenness condition in $\sfD 1$,  $\sfD 3$. 
 Alternatively, starting from  conditions  $\mathsf D3$ and $\mathsf D4$, we can conclude   that $\sum_{j } X_{ji}  K_{1jj}\in 2\bZ$. This means that the evenness conditions in  $\sfD$1,  $\sfD$2 can be relaxed as they can be derived from   $\sfD$3 and  $\sfD$4.
 
 The conditions above     applies to both bosonic and spin TQFTs.
 In   case that  the two sides are both bosonic, namely $K_1,K_2$ only have even diagonal entries, the above conditions reduce to that in  \cite{Delmastro:2019vnj}. The important differences in our generalized conditions  \eqref{dualityconditions} are that we impose some evenness conditions  in $\sfD$1 and $\sfD$2. Below we will show   a trick to relax these evenness condition by promoting  all TQFTs to spin TQFTs. 
 
  \PA{Evenness conditions. }
       To promote the a TQFT $\mathscr T$ to a spin TQFT $\widehat{ \mathscr T}$, we can just stack it with \sVec, namely  $\widehat{ \mathscr T}\cong\mathscr T\boxtimes  \sVec$.  \footnote{Stacking with $\sVec$  would also change the central charge, but we will ignore it for our purpose here.}  \footnote{Rigorously speaking, if $\mathscr T$ itself is spin, $\boxtimes$ should be more precisely denoted as $\boxtimes_\sVec$ in order to get rid of the transparent boson arising from the two transparent fermions of $\mathscr T$ and $\sVec$.  Formally $\cA \boxtimes_\sVec \cB=(\cA\boxtimes \cB)/ A$ where $A=(1_\cA,1_\cB)+(\psi_\cA,\psi_\cB) $. We will ignore the subscript $\sVec$ for simplicity.}
  Note that for spin TQFT $\mathscr T$, the stacking is trivial as  $ { \mathscr T}=\mathscr T\boxtimes  \sVec $.

 Let us assume that the  two TQFTs $\mathscr T_{K_1}$ and $ \mathscr T_{K_2}$  satisfy the following conditions for some integral matrices $R,R',P, P',X,X'$:
%
%
%
%
            \begin{subequations}\label{dualitys}
         \begin{align} 
    \widehat{ \sfD1}: &&      R  R'=1+K_1X~,  \\
  \widehat {  \sfD2}: &&       R'R =1+K_2X'  ~,   \\
   \widehat{ \sfD3}: &&          R^T K_1^{-1} R - K_{2}^{-1}= P~, \\
   \widehat   { \sfD4}: &&            R'^TK_{2}^{-1}R' -K_{1}^{-1}=P'~,  
                \end{align}
         \end{subequations}
           where we  have relaxed all the evenness conditions. We will prove that after promoting both theories to spin TQFTs,  the resulting spin TQFTs $\widehat{\mathscr T}_{K_1}$ and $ \widehat{\mathscr T}_{K_2}$ are dual to each other, namely they satisfy the duality conditions $\sfD1$, $\sfD2$, $\sfD3$, $\sfD4$   \eqref{dualityconditions} where all quantities  there are hatted.

At the level of $K$ matrix, the stacking is realized by  adding one  more diagonal entry 1 to $K$. \footnote{One can also add more than one diagonal entries consisting of both 1 and $-1$. Such a  generalization is straightforward.} We will denote the new $K$ matrix as $\widehat K$. As a result, we have         \be 
                 \widehat{    K_1} =\PBK{   K_1 &0 \\ 0 & 1}~, \qquad
                 \widehat{     K_2} =\PBK{   K_2 &0 \\ 0 & 1}~.
         \ee
         The resulting TQFTs can be written as  $\widehat{ \mathscr T}_K=\mathscr T_K\boxtimes   \sVec=\mathscr T_{\widehat K}$.
         Now let us construct two matrices 
               \be
         \widehat R=\PBK{ R& 0 \\ J & 1} ~, \qquad
                          \ee
      where $J =(J_1, J_2, \cdots )$ is an   integral vector.

         Then we can easily compute     
       \beqn
       \widehat   R^T \widehat{    K_1}^{-1} \widehat R
            &=&
           \PBK{ R^T K_1^{-1} R+ J^T J& J^T \\ J & 1}
           =         \PBK{     K_2^{-1} +P + J^T J& J^T \\ J & 1}
           = \widehat{    K}_2^{-1}   + \widehat P ~,
       \eeqn
where we used the condition $\widehat{\mathsf D3}$ above and  
         \be
         \widehat P = \PBK{ P + J^T J& J^T \\ J & 0}~.
         \ee
      The non-trivial diagonal entries of $\widehat P$ are   $   (  \widehat P)_{ii}=P_{ii}+J_i J_i$.  Obviously all the  diagonal entries of  $\widehat P$ become even if $ J_i=P_{ii}  \mod  2$.  Note $p^2\equiv p \mod 2 $ for $p\in \bZ$. A simple  particular choice is to take $ J_i=P_{ii} $.  As a result, the evenness conditions are satisfied, $ (  \widehat P)_{\hat i \hat i}\in 2 \bZ$ for all $\hat i$.

        Similarly, we can show that 
          \beqn
      \widehat{  R'}^T \widehat{    K_2}^{-1}    \widehat{  R'}
                    = \widehat{    K}_1^{-1}   + \widehat P' ~, \qquad
                     \widehat{  R'}= \PBK{  R'& 0 \\   J '& 1}    ~   , \qquad
                                   { \widehat P} ' = \PBK{ P' + J'^T J'& J'^T \\ J' & 0}~ , \qquad 
                                          \eeqn
where all the diagonal entries of ${ \widehat P}'$ are even   ${ \widehat P}'_{\hat i \hat i}\in 2\bZ$ if we choose $ J'_i =P'_{ii}\mod  2$.

           We can also show that
                \beqn 
         \widehat R \widehat {    R}'&=&  \PBK{ RR'& 0\\ J  R' + J' & 1}
          =\PBK{ 1+K_1X &0\\ J  R'+  J'  & 1} 
    =1+ \widehat K_1 \widehat X~,
         \eeqn 
         where  we used condition $\widehat{\mathsf D1}$ and
         \be
         \widehat X=\PBK{  X &  0  \\ J   R'+  J'  &0} ~.
         \ee
We can further show that 
      \be
  \sum_{\hat i} \widehat K _{1\; \hat i\hat i}   \widehat X_{\hat i \hat  j}\in 2\bZ~,
           \ee    
due to     the following identity
\be
 \sum_{\hat i} \widehat K _{1\; \hat i\hat i}   \widehat X_{\hat i \hat  j}  \stackrel{\mod 2\;\;\;\,}{=\!=\!=\!=}     \widehat  P'_{\hat i\hat i} - \sum_{\hat j} \widehat R'_{\hat j\hat i} \widehat P _{\hat j\hat j}  \stackrel{\mod 2\;\;\;\,}{=\!=\!=\!=}0~,
\ee
which can be analogously derived as  \eqref{evencds}. 

Similarly, one can show that
                \beqn 
         \widehat R' \widehat {    R}&=& 
    1+ \widehat K_2 \widehat X', \qquad
           \widehat X'=\PBK{  X '&  0  \\ J '  R +  J   &0} , \qquad
             \sum_{\hat i} \widehat K _{2\; \hat i\hat i}   \widehat X'_{\hat i \hat  j}\in 2\bZ~.
         \eeqn 
         Collecting all equations together, we find that  all the duality conditions $\sfD1$, $\sfD2$, $\sfD3$, $\sfD4$  (where all quantities are hatted) \eqref{dualityconditions}  are satisfied.
          
          To conclude,  if two  $K$-matrices satisfy the conditions   $\widehat{\sfD1}$, $\widehat{\sfD2}$, $\widehat{\sfD3}$, $\widehat{\sfD4}$ \eqref{dualitys}, then the corresponding TQFTs  $\mathscr T_{K_1}$ and $ \mathscr T_{K_2}$ 
 are dual to each other as \emph{spin} TQFTs, or more precisely, after stacking with \sVec, the  resulting spin TQFTs  $\widehat{\mathscr T_{K_1}}=\mathscr T_{\widehat K_1}= \mathscr T_{K_1}\boxtimes  \sVec$ and $\widehat{\mathscr T_{K_2}}=\mathscr T_{\widehat K_2}= \mathscr T_{K_2}\boxtimes  \sVec$  dual to each other and satisfy all the duality conditions $\sfD1$, $\sfD2$, $\sfD3$, $\sfD4 $ \eqref{dualityconditions}. Note that if $\mathscr T_{K_1}$ and $ \mathscr T_{K_2}$ are themselves spin TQFTs, then the evenness conditions can be relaxed automatically as 
spin TQFTs are invariant under the stacking with \sVec:  $ { \mathscr T}\cong\mathscr T\boxtimes \sVec$. 

In the following discussions, whenever we say  two TQFTs   
 are dual to each other as spin TQFTs or up to stacking with \sVec, it actually means that conditions   $\widehat{\sfD1}$, $\widehat{\sfD2}$, $\widehat{\sfD3}$, $\widehat{\sfD4}$ are all satisfied, and the duality holds after  promoting both of two TQFTs to spin TQFTs.

\subsection{Universal mass deformation: from mirror symmetry in  SCFT to duality in  TQFT}\label{mirrorduality}
Given the Abelian SCFT  $\cT_Q$ described above,  we can now  deform it by adding the universal mass term \eqref{unidef}. In general, it is difficult  to  keep track of the behavior of the theory after adding deformation due to the strongly coupled nature of  $\cT_Q$. Nevertheless, on can infer the candidate IR TQFT by looking at the UV gauge theory $T_Q$. See fig.~\ref{mirrorscfttqft}.
As discussed in\cite{KNBalasubramanian:2024uvi}, the resulting IR theory is given by some CS theory. We will review some of the important results which are crucial to establish the main claim of this paper later on.

\subsubsection{Proposal for IR TQFT
}
    
 One crucial consequence of universal mass deformation is that all the fermions would become massive. As a result, we  can integrate  out the fermions and get some  effective CS theory whose   CS level is specified by \cite{KNBalasubramanian:2024uvi}
     \be\label{Kab}
     K_{ab}=\frac12\sum_{i=1}^{N }2\sgn m_i\; Q_i^a Q_i^b = \sgn m \sum_{i=1}^{N } Q_i^a Q_i^b ~, \qquad
  \longrightarrow   \qquad
     K=\sgn m\;  QQ^T~,
     \ee
  where the factor of 2 comes from  2 fermions in each hypermultiplet \footnote{Note that each $\cN=4$ hypermultiplet contains two 3d $\cN=2$ chiral multiplets with opposite charges.}.  Also we use the fact that the masses of  all the fermions have the same sign under universal mass deformation, $\sgn m_i=\sgn m$. Without loss of generally, we assume they are positive. Note that here we don't consider contributions of fermions in vector multiplet as they are not charged U(1) gauge symmetry. 
  
  Similarly, the mirror theory $\cT_{\wt Q}$ would give rise to  
   an  effective CS theory with   CS level 
        \be
  \widetilde   K_{ab}=\frac12\sum_{j=1}^{ N }2\sgn\widetilde  m_j\; \widetilde Q_j^a \widetilde Q_j^b
   =   \sgn \wt m\sum_{j=1}^{ N }\widetilde Q_j^a\widetilde  Q_j^b  ~, \qquad
  \longrightarrow   \qquad
 \wt    K=\sgn \wt  m\;   \wt Q \wt Q^T~.
     \ee

 One interesting consequence of the proposal above is that the universal mass deformations of a mirror pair of two SCFTs    $\cT_Q, \cT_{\wt Q}$ would give rise to two TQFTs which are  dual to each other 
     \be\label{mirrorCS}
   U(1) ^k \text{ CS theory with   level } K=\sgn    m\;QQ^T
   \longleftrightarrow
  U(1) ^{N-k} \text{ CS theory  with   level } \widetilde K= \sgn \wt  m\;\wt Q \wt Q^T  ~,
     \ee
     if the sign of universal mass parameter is  flipped $\sgn\wt m=-\sgn m$. The flip of sign is crucial and can be understood from the algebra \eqref{QQalgebra}, where the automorphism of the algebra implementing the mirror symmetry also flips the sign of $m$. See figure~\ref{mirrorscfttqft} for illustration.
     Note that generally the duality between two TQFTs   above holds when only we regard the them as spin TQFTs. This means that   if we get a bosonic TQFT,   we need to add transparent fermion  $\psi$  which has topological spin $\theta(\psi)=-1$ and braids trivially with all the rest of anyons.  This point is also important for the familiar level-rank duality and has been  studied systematically in \cite{Hsin:2016blu}.

\begin{figure}
\[\begin{tikzcd}
	{T_Q} & {} &&&&&& {T_{\widetilde Q}} \\
	&& {\mathcal T_Q} &&& {\mathcal T_{\widetilde Q}} \\
	\\
	\\
	&& {\mathscr T_{K=QQ^T}} &&& {\mathscr T_{\widetilde K=-\widetilde Q\widetilde Q^T}}
	\arrow["{\text{RG flow}}", from=1-1, to=2-3]
	\arrow[dashed, from=1-1, to=5-3]
	\arrow["{\text{RG flow}}"', from=1-8, to=2-6]
	\arrow[dashed, from=1-8, to=5-6]
	\arrow["{\text{mirror symmetry}}", tail reversed, from=2-3, to=2-6]
	\arrow["{\delta { \mathcal L}=m  J}", from=2-3, to=5-3]
	\arrow["{\delta\tilde{ \mathcal L}=-m \tilde J}"', from=2-6, to=5-6]
	\arrow["{\text{duality}}", tail reversed, from=5-3, to=5-6]
\end{tikzcd}\]
 
\caption{Mirror symmetry and duality. Here  the vertical lines denote  the universal mass deformations.  $T_Q$ is a 3d $\cN=4$ supersymmetric gauge theory whose IR fixed point is  the SCFT $\cT_Q$. We also assume $m>0$ for simpliciy here.
 }\label{mirrorscfttqft}
\end{figure}
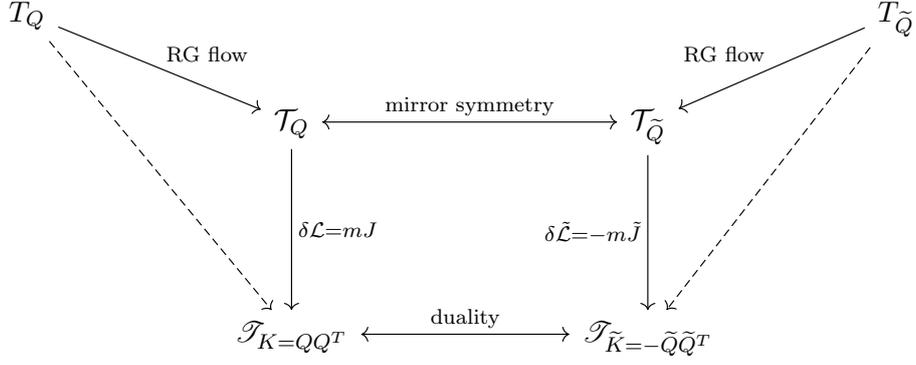

%
%
   \subsubsection{A simple   example  }  
     The simplest example is given by SQED, namely a single U(1) with $N$ hypermultiplets of charge $1$. The mirror is given by  a quiver gauge theory with $N-1$  U(1) gauge node. Such a mirror pair can be represented as
   
      \begin{figure}[h]\centering
\begin{tikzpicture}[node distance=2cm, auto, thick]

    \node (T) [  rectangle, draw  , minimum size=.61cm  ] at (-1, 0.1)  {$N$};
           \node (S) [circle,  draw, left of =T] {1};
        \draw[-] (S) -- (T);
        $\longleftrightarrow$
       
    \node (A) [rectangle,  draw, minimum size=.61cm] at (1, 0.1){1};
    \node (B) [circle, draw, right of=A ] {1};
    \node (C) [circle, draw, right of=B] {1};
    \node (D) [rectangle, draw, right of=C,minimum size=.61cm] {1};

    \draw[-] (A) -- (B);
    \draw[dashed ] (B) -- (C);
    \draw[- ] (C) -- (D);
\end{tikzpicture} 
\end{figure}
  \noindent where there are  $N-1$ circle   nodes on the RHS. If $N=2$, the two sides are actually the same and the corresponding theory is self-mirror symmetric. Their charge matrices for the two quivers are given by
     \be
     Q=\PBK{ 1& \cdots & 1}_{1\times N}~  \qquad\longleftrightarrow\qquad  
       \widetilde Q=
       \PBK{ 
   1&$-1$   &   &&   &\\
      &1   &$-1$   &  &   &\\
        &    &\vdots  &  &   &\\  
         &    &  & 1 & $-1$  &\\  
           &    &   &    &1   &$-1$\\  
           } _{(N-1)\times N}~.
     \ee
     One can check that indeed $\widetilde Q Q^T=0$.   After performing the universal mass deformations, the IR  theories  on  the two sides are CS theories with level matrices  \eqref{Kab}
     \footnote{The universal mass deformation is supposed to preserve all internal symmetries as well as their 't Hooft anomalies. Such a kind of  anomaly matching has indeed been verified between the SCFT arsing from SQED and   the $U(1) { }$ CS theory\cite{KNBalasubramanian:2024uvi}. This further justifies the proposal  in \eqref{Kab}.}
     \be
     K=-\PBK{N}  \qquad\longleftrightarrow  \qquad \widetilde K =C_{SU(N)}= 
     \PBK{ 
   2&$-1$   &   &&   &\\
      $-1$&2   &$-1$   &  &   &\\
        &    &\vdots  &  &   &\\  
         &    &  &    $-1$ &  2  & $-1$\\  
           &    &   &    & $-1$   &2\\  
           } _{(N-1)\times  (N-1)}~.
     \ee
     where we have chose $\sgn m =-\sgn \tilde m =1$, and $C_{SU(N)}$ denotes the Cartan matrix for   $SU(N)$.  
         Note the CS theories on the two sides have the same number of anyons $|\det K| =|\det\widetilde K|=N$. Moreover, the CS theory with level matrix   $C_{SU(N)}$ is nothing but the $SU(N)_1$ CS theory.  Therefore the duality  in \eqref{mirrorCS} is just \be
U(1)_{-N} \text{  CS theory } \leftrightarrow  
SU(N)_1 \text{ CS theory}=   U(1)^{ k-1} \text{  CS theory with   level } C_{SU(N)} ~,
\ee
which is the well-known level-rank duality. 
   \subsubsection{Duality from mirror symmetry}  
   Based on the proposal \eqref{Kab}, we will show that the  mirror pair of two SCFTs give rise to the dual pair of  TQFTs   \eqref{mirrorCS} after performing the universal mass deformation. 
   In order to prove the duality, we first need to derive serval matrix identities from \eqref{mirrorId} and \eqref{mirrorId2}, which can be then used to verify the duality conditions  \eqref{dualitys} in subsection \eqref{CSduality}.

Consider two matrices: 
   \be
   \Pi=Q^T (QQ^T)^{-1}Q~, \qquad
\wt   \Pi=\wt Q^T (\wt Q\wt Q^T)^{-1}\wt Q~.
   \ee
   Then it is easy to prove that 
   \be
   \Pi^2=\Pi~, \qquad
 \wt   \Pi^2=\wt \Pi~,\qquad
  \Pi \wt \Pi =\wt \Pi \Pi =0~.
   \ee
Therefore $(\Pi +\wt\Pi)^2 =\Pi +\wt\Pi$. This means that the matrix $\Pi +\wt\Pi$ can only have eigenvalues 1 or 0.  So $\Pi +\wt\Pi=U^{-1}D U$ where $D$ is diagonal matrix with entries 0 or 1. \footnote{Note that   $\Pi+\wt\Pi$ is real symmetric, and thus can be diagonalized all the time. }  Taking the trace, we find that 
     \be
\tr(\Pi +\wt\Pi)
    =        \tr(        (QQ^T)^{-1}Q Q^T)+   
    \tr(  (\widetilde Q\widetilde Q^T)^{-1}\widetilde Q   \widetilde      Q^T)
=k+(N-k)=N~.
     \ee
Therefore $\tr(\Pi +\wt\Pi)=\tr D =N$, which implies that $D$ is just the identity matrix. As a result, we conclude that 
\be
\Pi +\wt\Pi=Q^T (QQ^T)^{-1}Q+ \wt Q^T (\wt Q\wt Q^T)^{-1}\wt Q=\mathds{1}_N~.
\ee
This means $\Pi, \wt\Pi$ are projectors. 
We can then consider
\beqn
q(\Pi +\wt\Pi)q^T &=&
   (qQ^T)(QQ^T)^{-1}(Qq^T) +q \wt Q^T (\wt Q\wt Q^T)^{-1} ( q\wt Q^T)^T
\\   &=&
      (qQ^T)(QQ^T)^{-1}(Qq^T) + (\widetilde{Q}\widetilde{Q}^T)^{-1}
 = qq^T~,  \label{eq1}
\eeqn
  where we used \eqref{mirrorId2}. Similarly, we can also derive
      \be
( \wt  Q   \wt  q^T )( Q q^T)= \wt  Q  (\wt  q^T  Q) q^T= \wt  Q  ( \mathds{1}_N-\wt  Q^T q) q^T= \mathds{1}_k-(\wt Q   \wt Q^T )(q  q^T)    ~.
  \label{eq2}
       \ee

  We can define
       \be
       R= Qq^T~, \quad P= X' =qq^T~, \quad
           R'=  Q\wt q^T~, \quad  P'= X =-\wt q\wt q^T~, 
          \quad K_1=QQ^T, \quad   K_2=-\wt Q\wt Q^T~.
       \ee 
    Then the equations \eqref{eq1}  and  \eqref{eq2} above become
    \be 
        R^TK_1^{-1}R -K_2^{-1}=P~ , \qquad
      R'R = \mathds{1} +  K_2 X'~, \qquad 
    \ee    
   Similarly, one can also show
       \be 
         R'^T  K_2^{-1}  R '-  K_1^{-1}=  P'~, \qquad       R  R' = \mathds{1} +  K_1  X~. \qquad  
    \ee       
    It is easy to see that they four identities  above are exactly the duality conditions  $\widehat{\sfD1}$, $\widehat{\sfD2}$, $\widehat{\sfD3}$, $\widehat{\sfD4}$  \eqref{dualitys} in section~\ref{CSduality}.

     As a result, we succeed in showing the    the two TQFTs are dual to each other as spin TQFTs
       \be
 Q \wt Q^T=0: \qquad\qquad   \mathscr T_{K=QQ^T}
        \longleftrightarrow
             \mathscr T_{\wt K=-\wt Q\wt Q^T}~.
     \ee

      \section{Self-mirror symmetric   SCFT } \label{SMSCFT}
      
      In this section, we will construct the general family of self-mirror symmetric Abelian SCFTs. It turns out that the  charge matrix $Q$ of  self-mirror symmetric SCFT $\cT_Q$ satisfies an additional constraint \eqref{selfMirror}.   Self-mirror symmetric  SCFTs   exhibit numerous distinctive features. In particular, the CB and HB Hilbert series are identical, while the superconformal indices are invariant under the inversion of the fugacity for R-symmetry. The self-mirror symmetric SCFTs that we constructed indeed satisfy all these properties.  We also give some explicit examples of  self-mirror symmetric SCFTs, including a full classification of rank-1  self-mirror symmetric  Abelian SCFTs.

        \subsection{Hilbert series  for CB and HB }     
            A simple way to show the mirror symmetry between two theories is  computing their CB and HB Hilbert series.   Since   mirror symmetry exchanges the CB and HB, the  CB Hilbert series of a theory should be identical to the HB Hilbert series of its mirror, and vice versa. A more refined quantity is given by   the superconformal indices, whose specific limits  give rise to the CB and HB Hilbert series. A mirror pair of SCFTs have the same superconformal indices up to the inversion of    R-symmetry  fugacity. Given their importance, we will first review the general formulae for the  CB and HB Hilbert series of  abelian SCFT $\mathcal T_Q$ in this subsection, and  derive a  closed formula for  the superconformal index of $\mathcal T_Q$ in the next subsection. 
            
Consider Abelian gauge theory with charge matrix      $Q_a^i$   where   $a=1, \cdots, k$ labels the $k$ U(1) gauge groups,  and $i=1,\cdots N$ labels the $N$ hypermultiplets. The  CB and HB are parametrized by the vevs of the chiral operators, which form the chiral rings and can be described by the Hilbert series. The Hilbert series provides a simple way to enumerate the number chiral operators according to their scaling dimensions (and possibly charges for other symmetries).
     
     For Higgs branch, the corresponding Hilbert series can be computed  classically using the Molien integral for $F$-term equation. Explicitly, the Higgs branch Hilbert series  for a 3d $\cN=4$ SCFT  characterized by charge matrix $Q$ is given by \cite{Cremonesi:2013lqa}
            \be\label{HBHS}
      \HS_{\cT_Q}^{\HB}(q) =(1-q )^k \Bigg( \prod_{a=1}^k \oint_{|z_a|=1}\frac{dz_a}{2\pi i z_a} \Bigg)
      \frac{1}{ \prod\limits_{i=1}^N \Bigg[1-q^{\frac12} \Big( \prod\limits_{a=1}^k z_a^{Q_a^i}  + \prod\limits_{a=1}^k z_a^{-Q_a^i} \Big)
      +q\Bigg] }~.
      \ee

     For Coulomb  branch, the corresponding chiral operators are the dressed monopoles.   The corresponding Hilbert series can be computed using the  monopole formula. Explicitly, the Coulomb branch Hilbert series  for a  3d $\cN=4$ SCFT  characterized by charge matrix $Q$ is given by \cite{Cremonesi:2013lqa}

\be\label{CBHS}
      \HS_{\cT_Q}^{\CB}(q) =\frac{1}{(1-q)^{ k}}\sum_{\bm \fkm \in \bZ^{ k}} 
      q^{\frac12\sum_{i=1}^N \Big| \sum_{a=1}^{k}
  Q_i^a \fkm_a
\Big|}~.
\ee   
            \subsection{Superconformal index}   
            The Coulomb  branch and Higgs branch Hilbert series only capture  the chiral operators which are $\frac12$-BPS. To get refined information for less protected operators, we can consider superconformal indices.

            In general, the 3d $\cN=4$ superconformal indices are defined as  \cite{Imamura:2011su,Kapustin:2011jm,Li:2023ffx}
\be
\cI =\Tr (-1)^{2J_3} x^{\Delta+J_3} t^{2(R_H-R_C)}\prod_i \mu_i ^{F_i} ~,
\ee
where  $\Delta$ is the scaling dimension, $J_3$ is the spin for the $SO(3)\sim SU(2)$ Lorentz symmetry, $R_H,R_C$ are the charges for the $SU(2)_H\times SU(2)_C$  $R$-symmetries,
\footnote{The spins $J_3,R_H,R_C$ are half-integer quantized, and are related to the Dynkin labels    for multiplets in \ref{universalmassdef} via $2J_3=j,2R_H=R,2R_C=R'$.}
 and $x,t$ are the corresponding fugacities for these symmetries. We also include the fugacities and charges for the global flavor symmetry, denoted as $\mu_i$ and $F_i$ respectively. 
The superconformal indices get  contributions from local operators satisfying $\Delta-R_H-R_C-J_3=0$ which are generally $\frac18$-BPS.

For 3d $\cN=4$ SCFTs coming  from SUSY gauge theories, one can regard the superconformal indices as the partition functions on $S^3\times S^1$ and then compute the indices using localization. The general strucuture is as follows:
 \be
 \cI  =\sum_\fkm \frac{1}{|W_\fkm|}\oint \frac{d\bm z}{2\pi i \bm z}
 \cZ_\text{cl}\cZ _\text{vec}\cZ_\text{hyp}~,
 \ee 
 where $\cZ_\text{cl}$ is the  classical contribution, while     $ \cZ _\text{vec}$ and $\cZ_\text{hyp}$ are the 1-loop contribution from vector and hyper multiplets, respectively. 
 Here $\bm z$ and $\bm \fkm$ are the fugacities and   magnetic fluxes for the gauge symmetry. $|W_{\bm\fkm}|$ is the order of the Weyl group of  the residual gauge symmetry group  in the monopole background with flux $\bm\fkm$.
 There are similar fugacities and     fluxes  for  each flavor symmetry which are not shown explicitly. Note that in 3d $\cN=4$, we have not only flavor symmetry acting on the hypermultiplets, but also the topological symmetries acting on monopoles.

 In this paper, we   consider the 3d $\cN=4$ SCFTs arsing   as the IR fixed points of  Abelian  gauge theories. As we discussed before, such SCFTs are fully specified by a charge matrix $Q_i^a$. For simplicity, we turn off all the fugacities and fluxes for all the flavor symmetries. In this case, using the formula in e.g. \cite{Li:2023ffx},  one can show the the index takes the following form:
       \beqn\label{AbSCI}
    \cI&=&
        ( Z_V  )^{\text{rank }G}
        \sum_{\bm\fkm \in \bZ^{\text{rank }G} } \Big( \prod_{a=1}^ {\text{rank }G} \oint_{|z_a|=1}\frac{dz_a}{2\pi i z_a} \Big)
         \prod_ {\rho\in \bm R} Z_H\Big(\rho(\fkm), \bm z^{\rho} \Big)~,
     \eeqn
     where $\rho$ is the matter representation and
         \beqn
 Z_V & =&
   \frac{(t^2 x, x^2)_\infty}{(t^{-2} x, x^2)_\infty} ~,  \qquad\qquad (a,y) =\prod_{j=0}^\infty (1-a y^j)~,  \label{ZV}
   \\            
  Z_H(m,z)& =& \left(\frac{x}{t^2}\right)^{\frac{1}{2}| m| } \prod_{A=\pm 1} \prod_{B=\pm 1}
  \left( (-1)^m x^{|m|+1}    z^A \left(\frac{\sqrt{x}}{t}\right)^B;x^2\right)_{\infty }^B~.
  \label{ZH}
       \eeqn
            
            Obviously, we have
            
            \be\label{ZHmz}
              Z_H(m,z)=  Z_H(m,1/z)=  Z_H(-m,z)~.
            \ee

         For a hypermultiplet with charge $Q^a$ under the $a$-th U(1) gauge group, we have
              \be
      \rho(\bm\fkm)=\sum_a Q^a \fkm_a~, \qquad
      \bm z^\rho=\prod_a z_a^{Q^a}~.
      \ee
      
      As a result, we get the superconformal index of SCFT $\cT_Q$ with rank $k$ and $N$ hypermultiplets:
               \beqn\label{IindexQ}
    \cI _{\cT_Q} &=&
       \Bigg(  \frac{(t^2 x, x^2)_\infty}{(t^{-2} x, x^2)_\infty} \Bigg)^k \times
        \sum_{\bm\fkm \in \bZ^k } \Big( \prod_{a=1}^k \oint_{|z_a|=1}\frac{dz_a}{2\pi i z_a} \Big)
         \prod_{i=1}^N Z_H\Big( \sum_{a=1}^k  Q_i^a \fkm_a\, ,\; \prod_{a=1}^k z_a^{Q_i^a} \Big)~.
         \eeqn
       
         We can further consider specific limits of the (unflavored and unfluxed) index which are related to the moduli space of the underlying SCFT. For this purpose, we introduce another two fugacities $q_h, q_c$ via 
      \be \label{qchtx}
      q_h q_c=x^2~, \qquad q_h/q_c=t^4~, \qquad\qquad q_h = {x }   t^2~,  \qquad q_c = {x }  t^{-2}~.
      \ee
      
      The   Coulomb/Higgs index is  then defined as the following limit  of the general $\cN=4$ superconformal index: 
      \begin{itemize}
     \item Coulomb index: $q_h\to 0$, $q_c$ fixed
      
   \item  Higgs index: $q_c\to 0$, $q_h$ fixed
        \end{itemize}
     In these limits, one expects that the   Coulomb/Higgs index  coincides with the Hilbert series of the  Coulomb/Higgs branch.        
          Let us show  this explicitly.  Using the fugacity map \eqref{qchtx}, we can express \eqref{ZV}\eqref{ZH} in terms of $q_h,q_c$, and then take the Coulomb/Higgs limit:
      \beqn
 Z_V & =&
\frac{(q_h, q_hq_c)_\infty}{(q_c, q_hq_c)_\infty}
   =     \begin{cases}
          1-q_h~, & q_c \to 0, \; q_h \text{ fixed}~,\\   \\[-2.008ex] 
  \frac{1}{  1-q_c }~,& q_h\to 0, \; q_c \text{ fixed}~,
          \end{cases} 
       \eeqn
       and   \beqn
         Z_H(m,z)&=&q_c^{\frac{| m| }{2}} \prod_{A=\pm 1} \prod_{B=\pm 1}
          \left((-1)^m z^A q_c^{\frac{1}{2} (B+| m| +1)} q_h^{\frac{| m| +1}{2}};q_c q_h\right)_{\infty }^B
          \\&=&
          \begin{cases}
          \frac{\delta_{m,0}}{1-  \sqrt{ {q_h}} (z+\frac 1 z)  + {q_h} } ~,& q_c\to 0, \; q_h \text{ fixed} ~,\\   \\[-2.008ex] 
          {q_c}^{\frac{| m| }{2}} ~, & q_h\to 0, \; q_c \text{ fixed}~.
          \end{cases}  
   \eeqn
   
    With these equations, it is obvious that that \eqref{IindexQ} reproduces the expressions   \eqref{CBHS}\eqref{HBHS} in the Coulomb/Higgs limit, respectively. Namely, we have
    \be\label{HSind}
        \HS_{\cT_Q}^{\HB}(q) =\cI_{\cT_Q}(q_c=0, q_h=q)~, \qquad
                \HS_{\cT_Q}^{\CB}(q) =\cI_{\cT_Q}(q_c=q, q_h=0)~, \qquad
    \ee
    where the index $\cI_{\cT_Q}$ is now  a function of $q_c,q_h$ via \eqref{qchtx}. Therefore, we have succeeded in reproducing the expressions of CB and HB Hilbert series, thus justifying that our general formula of superconformal index \eqref{IindexQ} is indeed correct.

        \subsection{Self-mirror symmetry}
    Under mirror symmetry, the HB and CB are exchanged. Consider two SCFTs $\cT$ and $   \cT^\vee$ which are the mirror of each other, then their Hilbert series and indices are related 
    \be
            \HS^{\HB}_\cT (q)=        \HS ^{\CB}_{ \cT^\vee}(q)~, \qquad
     \HS ^{\CB}_\cT (q)=        \HS ^{\HB}_{ \cT^\vee}(q)~,\qquad
       \cI_\cT (x,t)=     \cI_{ \cT^\vee} (x,t^{-1})~.
    \ee
    
    It has been proved mathematically in \cite{Cremonesi:2013lqa} that for two SCFTs $\cT_Q $ and $\cT_{\wt Q}$  
    \be\label{HSCBHBTQ}
    Q\wt Q^T=0~,\qquad \Longrightarrow   \qquad         \HS^{\HB}_{\cT_Q} (q)=        \HS ^{\CB}_{ \cT_{\wt Q}}(q)~, \qquad
     \HS ^{\CB}_{\cT_Q} (q)=        \HS ^{\HB}_{{ \cT_{\wt Q}}}(q)~, \qquad
    \ee
    for matrices $Q_i^a\in \bZ^{k\times N},\wt Q_j^b\in \bZ^{(N-k)\times N}$  satisfying $\sum_{i=1}^N Q_i^a \wt Q_i^b=0$,   $\forall a=1, \cdots, k , b= 1, \cdots, N-k$. This provides strong hints that $\cT_Q $ and $\cT_{\wt Q}$  are related by mirror symmetry, and we expect that    \footnote{It is interesting to prove this formula directly at the level of mathematics, but we will not try here. }
    \be
         \cI_{\cT_Q} (x,t)=     \cI_{{ \cT}_{\wt Q} }(x,t^{-1})~.
    \ee

    We are particularly interested in the self-mirror case, namely the case $  \cT^\vee=\cT$. In this case, we easily see that the HB and CB are the same, sharing the same       identical Hilbert series, and the  superconformal index should be  invariant under the  exchange of $q_c$ and $q_h$:     
        \be\label{mirrorind}
            \HS^{\HB}_\cT (q)=        \HS ^{\CB}_{ \cT}(q)~, \qquad 
       \cI_\cT (x,t)=     \cI_{ \cT} (x,t^{-1})~.
    \ee

Since mirror symmetry maps a theory with rank $k$ to a theory with rank $N-k$,  self-mirror symmetry would only be possible if the two ranks are the same, namely  $N=2k$.

As we discussed before, $\cT_Q $ and $\cT_{\wt Q}$ are related by mirror symmetry if
\be\label{QQT}
Q\wt Q^T=0~.
\ee
So for a theory $\cT_Q$ to be  self-mirror symmetric, naively we should require $\wt Q=Q$, and thus $QQ^T=0$. However, $QQ^T=0$ can never hold  for nonzero matrix $Q$, as its diagonal entries   $(QQ^T)_{aa}=\sum_{i}(Q_i^a)^2$ vanish  if and only if $Q^a_i=0$ for $\forall a,i$.

In order to find the self-mirror symmetric theory, one should use the fact that map from $Q$ to $\cT_{  Q}$ is not injective. Namely, it could happen that two different charge matrix $Q,Q'$ may describe the same SCFT $\cT_Q=\cT_{Q'}$. This kind of redundancy is already visible from \eqref{Qtsf} where a change of basis of vector  fields  leads to the same SCFT. For hypermultiplets, the change of   basis   is generally not allowed. Nevertheless, there are some freedoms that we can use to change $Q$ without modifying the SCFT $\cT_Q$. 

First of all, we can obviously relabel the set of hypermultiplets.  In term of charge matrix, the reshuffling of hypermultiplets  corresponds to 
\be
Q\mapsto QM~,
\ee
where $M$ is a permutation matrix.  
Secondly, we note that a  $\cN=4$ hypermultiplet with charge $Q$ consists of two $\cN=2$  chiral multiplets with charge $Q$ and $-Q$, respectively. This means a hypermultiplet with charge $Q$  is the same as a hypermultiplet with charge $-Q$. In term of charge matrix, the flipping of signs  corresponds to 
\be
Q\mapsto Q\Lambda~,
\ee
where $\Lambda$ is a   diagonal  matrix whose  diagonal entries are $\pm 1$.  

Combining the two operations  above together, we find the general action on a charge matrix  is given by  
\be
Q\mapsto Q\Omega~,
\ee
where $\Omega=\Lambda M =M \Lambda'$. Obviously, the set of $\Omega$ forms a group, called  hyperoctahedral group $\cB_N$. It consists of all signed permutations of $N$ elements.
 
 The matrix $\Omega$ is unimodular $|\det \Omega|=1$ and satisfies 
 \be
 \Omega^T\Omega=M^T\Lambda^T   \Lambda M=M^T  M =1~.
 \ee
Combining    the left multiplication by unimodular matrix, and the right-multiplication by   signed permutation matrix, we get the general automorphism of charge matrix 
\be\label{Qauto}
Q\mapsto  AQ\Omega~, \qquad A\in \GL(N, \bZ)~, \qquad \Omega\in \cB_N~.
\ee
One can verify that this kind of action leaves the Hilbert series and superconformal indices invariant. For example, if we look at the Coulomb branch Hilbert series in  \eqref{CBHS}, the left multiplication is inessential due to the sum over all integral flux, while flipping the signs of charge for each hypermultiplet or relabelling the 
the hypermultiplets also leave the Hilbert series untouched. The same kind of statements can be proved for  Higgs branch Hilbert series in  \eqref{HBHS}, and  superconformal index \eqref{IindexQ} where one can use the equation \eqref{ZHmz}.

To get a self-mirror symmetric SCFT, we need to find   $\wt Q=AQ\Omega$ such that \eqref{QQT}
\be\label{qqTA}
Q\wt Q^T=Q \Omega^T Q^T A^T=0~.
\ee
Since $A$ is unimodular, we conclude that the condition of self-mirror symmetry is  
\be\label{selfMirror}
Q \Omega   Q^T=Q \Omega  ^T Q^T=0~, \qquad \Omega\in \cB_N~, \qquad Q \text{ is primitive}~.
\ee
With this condition, one can then prove that  
\be\label{HSSM}
     \HS^{\HB}_{\cT_Q}(q)=     \HS^{\CB}_{\cT_{  Q}}(q)~, \qquad 
     \ee
   which follows from \eqref{HSCBHBTQ} and \eqref{qqTA}. Furthermore, from \eqref{HSind} we also expect
\be\label{IndSM}
          \cI_{\cT_Q}(x,t)=    \cI_{\cT_{  Q}}(x,1/t)~.
\ee

 We can  now try to find solutions to \eqref{selfMirror}.  Due to the redundancy  \eqref{Qauto} discussed before, the physically inequivalent solutions  can be represented as
\be
\cS=\GL(k, \bZ)\backslash  \big\{Q\in \bZ^{k\times N} | Q \text{ is primitive, and } \exists \Omega\in \cB_N   \text{ such that } Q \Omega   Q^T=0 
\big\}/\cB_N~.
\ee
If we find a solution, $Q \Omega   Q^T=0$, then there are automatically many other solutions $Q' \Omega ' Q'^T=0$ for $Q'=Q\widehat\Omega,\Omega ' =\widehat\Omega^T\Omega ' \widehat\Omega$, which is conjugate to $\Omega$. So we just need to focus on the conjugacy class of $\cB_N$ and find solutions for each conjugacy class 
\be
\cS= \bigcup_ {[\Omega]\in \cB_N} \cS_{[\Omega]}~.
\ee 
Note the conjugate class of $\cB_N$ is given by a pair of partitions $(\lambda, \mu)$ such that $|\lambda|+|\mu|=N$.  
 
 \subsection{Examples}\label{selfmirrorexample}

Now  we can show some explicit examples.  Let us start with the simplest case that $k=1$. Suppose the matrix $Q,\Omega$ have the following  form
\be
 Q=(m,n )~, \qquad
 \Omega= \PBK{ 0& -1 \\ 1 & 0}~ ,
 \ee
 then the  self-mirror symmetry condition is
 \be
 Q\Omega Q^T 
 =mn-nm=0~,
 \ee 
 which is trivially satisfied because $m,n$ are just numbers.  In this case $\cB_2$ is 	dihedral group $D_4$.  The other condition we need to impose is that $Q$ should be primitive, which leads to the condition $\gcd(m,n)=1$.    
Therefore, the self-mirror symmetric Abelian SCFT at rank 1 is $\cT_{Q=(m,n)}$,   classified by two coprime integers $m,n$.   
The simplest example is  $Q=(1,1)$, which gives rise to SQED with 2 flavors, and it is also known as $T[SU(2)]$ SCFT.  
Then using  \eqref{HBHS}\eqref{CBHS}\eqref{AbSCI}, one can   easily compute  the superconformal index and Hilbert series  
    \beqn
  \cI&=&
  1+ {3 \left(t^2+ t^{-2}\right) x} + \left(5 t^4+ {5}{t^{-4}}-7\right) x^2 + {\left(7 t^{6}-4 t^2-4 t^{ -2}+7t^{-6}\right) x^3} +\cdots ~,\qquad\qquad 
\\
\HS^\text{CB}=\HS^\text{HB}&=&1 + 3 q + 5 q^2 + 7 q^3 + 9 q^4 + 11 q^5+\cdots~.
\eeqn
It is easy to see that  that \eqref{IndSM} and \eqref{HSSM} are indeed satisfied. 

 Similarly for $Q=(2,3)$,  we have
 \beqn
\cI&=&1 + \frac{\left(t^4 + 1\right) x}{t^2} + \frac{\left(t^8 - 3 t^4 + 1\right) x^2}{t^4} + \frac{\left(t^{12} + 1\right) x^3}{t^6} + \left(2 t^5 + \frac{2}{t^5}\right) x^{5/2}+\cdots~,
\\
\HS^\text{CB}=\HS^\text{HB}&=&
1 + q + q^2 + 2 q^{5/2} + q^3 + 2 q^{7/2} + q^4 + 2 q^{9/2} + 3 q^5 + 2 q^{11/2} + 3 q^6 +
+\cdots~.\qquad\qquad\qquad
 \eeqn
 
 For $k=2$, we can assume the following form of $\Omega$
 \be
Q= \left(
\begin{array}{cccc}
 a & b & c & d \\
 e & f & g & h \\
\end{array}
\right)~,
\quad
\Omega=\left(
\begin{array}{cccc}
 0 & 0 & 0 & 1 \\
 0 & 0 & 1 & 0 \\
 0 & -1 & 0 & 0 \\
 -1 & 0 & 0 & 0 \\
\end{array}
\right)~,
\quad
Q\Omega Q^T=(d  e + c  f - b  g - a  h)
\left(
\begin{array}{cc}
 0 & -1 \\
 1 & 0 \\
\end{array}
\right)~.
 \ee
Therefore  $ Q\Omega Q^T=0$ if only if $a h+b g-c f-d e=0$. One solution is given by
\be\label{rk2Q}
Q=\left(
\begin{array}{cccc}
 1 & 1 & 1 & 1 \\
 1 & 0 & 1 & 0 \\
\end{array}
\right)~,
\ee
whose superconformal index and Hilbert series are given by
\beqn
\cI&=&1 + \frac{6 (1 + t^4)}{t^2} x + \left(4 + \frac{19}{t^4} + 19 t^4\right) x^2 + 
\frac{4 (11 - 5 t^4 - 5 t^8 + 11 t^{12})}{t^6} x^3
\nonumber
\\ &&
+ \left(65 + \frac{85}{t^8} - \frac{60}{t^4} - 60 t^4 + 85 t^8\right) x^4
+ \frac{2 (73 - 58 t^4 + 32 t^8 + 32 t^{12} - 58 t^{16} + 73 t^{20})}{t^{10}} x^5+\cdots~, \qquad \quad
\eeqn
and
\begin{equation*}
\HS=1 + 6 q + 19 q^2 + 44 q^3 + 85 q^4 + 146 q^5 + 231 q^6 + 344 q^7+\cdots~.
\end{equation*}

One can try to solve  the constraints \eqref{selfMirror} on charge matrix in general, but we will not attempt  here.  Instead let us  take a different perspective and consider a special class of self-mirror symmetric SCFT which can be engineered from D3-D5-NS5 brane  in type IIB string theory \cite{Hanany:1996ie}. The corresponding 3d theory is given by a linear quiver theory, usually denoted as  $T_\rho^\sigma[SU(M)]$ \cite{Gaiotto:2008ak}, where  $\rho, \sigma $ are the partitions of $M$, namely $\rho=[\rho_1, \rho_2, \cdots]$ subject to $\sum_i \rho_i=M$, and similarly for $\sigma$. We will denote the transpose of $\sigma$ as $  \sigma^T=[ \sigma^T_1, \sigma^T_2, \cdots]$. The linear quiver theory has gauge nodes   $U(N_i)$ connected linearly by bi-fundamental hypermultiplets, together with $F_i$ fundamental hypermultiplets attached to each gauge node $U(N_i)$. See  Fig.~\ref{quiver}.  The data of $N_i, F_i$ are related to the partitions $\rho, \sigma$ as follows:   
\be\label{FNrhosigma}
F_i = \sigma^T_i - \sigma^T_{i+1}~ , \qquad N_i =\sum_{j>i}\rho_j-\sum_{j>i} \sigma^T_j  ~,
\ee
or equivalently
\be\label{FNrhosigma2}
 \sigma^T_i=\sum_{j\ge i }F_j~, \qquad
\rho_i = \sigma^T_i +N_{i-1}-N_i~.
\ee
%
%
The number of gauge nodes, namely the number of non-zero $N_i$, is denoted by $k$. Then one can show that 
\be
\sigma=[k^{F_k}, (k-1)^{F_{k-1}}, \cdots, 1^{F_1}]~,
\ee
and consequently
\be
M=\sum_i iF_i=\sum_i i ( \sigma^T_i - \sigma^T_{i+1})=\sum_i   \sigma^T_i =\sum_i   \sigma _i  ~.
\ee

 In order to get an interacting SCFT in the IR, the quiver needs to be good such that $N_{i+1}+N_{i-1}+F_i\ge 2 N_i$ \cite{Gaiotto:2008ak}. This is guaranteed if    $\rho$ and $\sigma$ are \emph{ordered} partitions of $N$, namely  $\rho_1\ge \rho_2\ge\cdots$, and similarly for $\sigma$.  \footnote{Indeed, from \eqref{FNrhosigma} and \eqref{FNrhosigma2}, one can easily see that $\rho_i-\rho_{i +1}=\sigma^T_i - \sigma^T_{i+1} +N_{i-1}+N_{i+1}-2N_i=F_i+N_{i-1}+N_{i+1}-2N_i\ge 0$. In addition,  we should also impose the condition that $F_i$  and $N_i$ are all non-negative, which yields the further constraints  that  $\sigma$ should  also  be an ordered partition and  $\rho\le \sigma^T$ (namely $\sum_{j\le i}\rho_j \le \sum_{j \le i} \sigma^T_j  $ for all $i$ ). }

 \begin{figure}[h]\centering
\begin{tikzpicture}[node distance=2.2cm, auto, thick]
    \node (A) [circle, draw,  minimum size=1.2991cm] {$N_1$};
    \node (B) [circle, draw, right of=A ,  minimum size=1.2991cm] {$N_2$};
    \node (C) [circle, draw, right of=B,  minimum size=1.2991cm] {$N_3$};
    \node (D) [circle, draw, right of=C,  minimum size=1.2991cm] {$N_{k-1}$};
    \node (E) [circle, draw, right of=D,  minimum size=1.2991cm] { $N_k$};

    \node (A1) [rectangle, draw, above of=A,  minimum size=.991cm] {$F_1$};
    \node (B1) [rectangle, draw, above of=B,  minimum size=.991cm] {$F_2$};
    \node (C1) [rectangle, draw, above of=C,  minimum size=.991cm] {$F_3$};
    \node (D1) [rectangle, draw, above of=D,  minimum size=.991cm] {$F_{k-1}$};
    \node (E1) [rectangle, draw, above of=E,  minimum size=.991cm] {$F_k$};

    \draw[-] (A) -- (B);
    \draw[- ] (B) -- (C);
    \draw[dashed ] (C) -- (D);
    \draw[- ] (D) -- (E);
    \draw[- ] (A1) -- (A);
    \draw[- ] (B1) -- (B);
    \draw[- ] (C1) -- (C);
    \draw[- ] (D1) -- (D);
    \draw[- ] (E1) -- (E);
\end{tikzpicture}
\caption{Linear quiver gauge theory $T_\rho^\sigma[SU(N)]$. Each circle with label $N_i$ represents  a $U(N_i)$ gauge node, and there are $k$ gauge nodes in total. Each square node with  label $F_i$ standards for $F_i$ fundamental hypermultiplets. Each   line  connecting two   nodes represents a bi-fudamental/fundamental hypermultiplet. }
\label{quiver}
\end{figure}
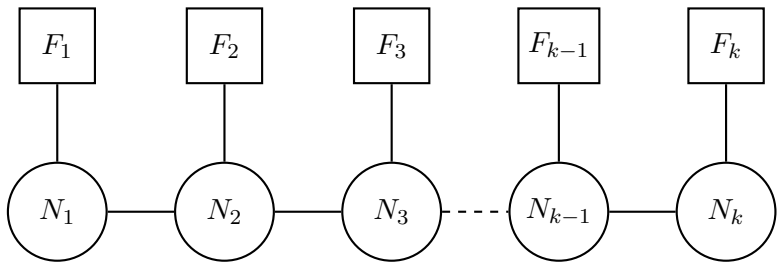

The mirror symmetry acts by exchanging $\rho$ and $\sigma$:
\be
T_\rho^\sigma[SU(M)] \longleftrightarrow (T_\rho^\sigma[SU(M)])^\vee= T^\rho_\sigma[SU(M)]~.
\ee

In this paper, we are interested in the self-mirror theory, so we focus on the case    $\rho=\sigma$. 

 If we further require that all the gauge nodes are Abelian, namely $N_i=1$, we get a stronger condition that 
\be
\sum_{j>i}\sigma_j-\sum_{j>i} \sigma^T_j=\Theta(i\ge 1) \Theta( i\le k)~,
\ee
where $\Theta(x)$ equals to $1$ if the argument  $x$ is true and 0 otherwise.   The corresponding theory is given by $k$  U(1) gauge nodes and $2k$ hypermultiplets with charge $\pm 1$. 
Solving these constraints, we easily see that 
\be
\rho=\sigma=[\lambda, 1]=[\lambda_1~, \cdots, \lambda_k, 1]~,  \qquad   \sigma^T=[\lambda_1+1, \cdots, \lambda_k]~,
\ee
where $ \lambda=[\lambda_1, \cdots, \lambda_k]$ is invariant under transpose, namely $  \lambda^T=\lambda$. In particular, we have $\lambda_1=k$. So $\sum_i F_i= \sigma^T=k+1$. Therefore, the total number of hypermultiplet is given by $k-1+\sum_i F_i =2k$, which is a necessary condition for self-mirror symmetry. 

To conclude, the self-mirror symmetric Abelian linear quiver gauge theory constructed from D3-D5-NS5 brane system is given by  $T_\sigma^\sigma[SU(M)]$ with  $\sigma=[\lambda, 1]$ where $\lambda=  \lambda^T$ is transpose invariant. This gives  infinitely many theories. For example, we can take $\lambda=[3 ,1,1]$, then $\sigma=[3,1,1,1]$. The resulting quiver theory has $\bm N=(1,1,1)$ and $\bm F=(3,0,1)$. One can verify that the charge matrix satisfies the self-mirror condition:
\be
Q=
\left(
\begin{array}{cccccc}
 1 & 0 & 0 & 1 & 1 & 1 \\
 -1 & 1 & 0 & 0 & 0 & 0 \\
 0 & -1 & 1 & 0 & 0 & 0 \\
\end{array}
\right)~,
 \qquad 
 \Omega=\left(
\begin{array}{cccccc}
 0 & 0 & 0 & 1 & 0 & 0 \\
 0 & 0 & 0 & 0 & 1 & 0 \\
 0 & 0 & 0 & 0 & 0 & 1 \\
 -1 & 0 & 0 & 0 & 0 & 0 \\
 0 & -1 & 0 & 0 & 0 & 0 \\
 0 & 0 & -1 & 0 & 0 & 0 \\
\end{array}
\right)~,
\qquad
Q\Omega Q^T=0~.
\ee 
 
 To be more concrete, we consider    a specific type of theory with $\sigma=\rho=[k,k,k-1,k-2, \cdots, 2,1]$ whose corresponding quiver  is  shown in figure~\ref{selfquiver}.  We see that there is only one balanced node, so the topological symmetry is enhanced from $\fku(1)^{k }$ to $\fks\fku(2)\times \fku(1)^{k-1}$.  Similarly, the hypermultiplet gives rise to flavor symmetry $\fks\fku(2)\times \fku(1)^{k-1}$, which coincides with the topological symmetry as required by self-mirror symmetry.
 \begin{figure}[h]\centering
\begin{tikzpicture}[node distance=2cm, auto, thick]
    \node (A) [circle, draw] {1};
    \node (B) [circle, draw, right of=A ] {1};
    \node (C) [circle, draw, right of=B] {1};
    \node (D) [circle, draw, right of=C] {1};
    \node (E) [circle, draw, right of=D] {1};

    \node (A1) [rectangle, draw, above of=A,  minimum size=.61cm] {1};
    \node (B1) [rectangle, draw, above of=B,  minimum size=.61cm] {1};
    \node (C1) [rectangle, draw, above of=C,  minimum size=.61cm] {1};
    \node (D1) [rectangle, draw, above of=D,  minimum size=.61cm] {1};
    \node (E1) [rectangle, draw, above of=E,  minimum size=.61cm] {2};

    \draw[-] (A) -- (B);
    \draw[- ] (B) -- (C);
    \draw[dashed ] (C) -- (D);
    \draw[- ] (D) -- (E);
    \draw[- ] (A1) -- (A);
    \draw[- ] (B1) -- (B);
    \draw[- ] (C1) -- (C);
    \draw[- ] (D1) -- (D);
    \draw[- ] (E1) -- (E);
\end{tikzpicture}
\caption{Quiver for a special type of self-mirror SCFT.  All hypermultiplets have charge $\pm1$.  }
\label{selfquiver}
\end{figure}
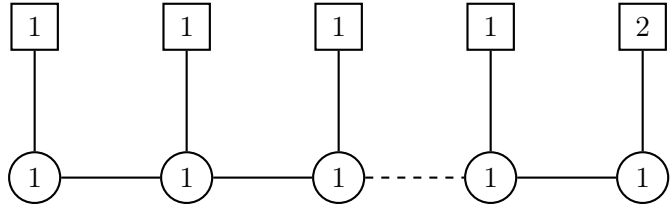

 The charge matrix is given by 
 \be
 Q= \PBK{ \mathds{1}_k &\quad  \mathds{1}_k -J_k }~,
 \ee
 where the matrix   entries are $(J_k)_{ij}=\delta_{i,j+1}$.   We can   choose  
  \be
  \Omega =\PBK{0 & \omega_k \\ -\omega_k & 0}~,
 \ee
 where $\omega_k$ is the  exchange matrix whose non-zero entires are  1 on the   antidiagonal. More explicitly, the entries are
 $\Omega_{ij}= \delta_{i+j, 2k+1}\sgn(j-i )$.  
 One can easily prove that 
 \be
 Q\Omega Q^T=-\omega_k +J_k  \omega _k+ \omega_k-\omega_k J_k^T=  J_k\omega_k- \omega_k J_k^T 
 = \omega_k(\omega_k J_k\omega_k-   J_k^T)=0~.
 \ee

 For example, if $k=3$, we have
 \be\label{rk3Q}
 Q=\left(
\begin{array}{cccccc}
 1 & 0 & 0 & 1 & 0 & 0 \\
 0 & 1 & 0 & -1 & 1 & 0 \\
 0 & 0 & 1 & 0 & -1 & 1 \\
\end{array}
\right)~,
\qquad
   \Omega =\left(
\begin{array}{cccccc}
 0 & 0 & 0 & 0 & 0 & 1 \\
 0 & 0 & 0 & 0 & 1 & 0 \\
 0 & 0 & 0 & 1 & 0 & 0 \\
 0 & 0 & -1 & 0 & 0 & 0 \\
 0 & -1 & 0 & 0 & 0 & 0 \\
 -1 & 0 & 0 & 0 & 0 & 0 \\
\end{array}
\right)~.
 \ee
  
Actually, we can generalize the above theory by modifying the charge matrix as follows
 \be
 Q= \PBK{ \mathds{1}_k &\quad  \mathds{1}_k -sJ_k }, \qquad s\in \bZ~,
 \ee
where $s$ is an arbitrary integer.  One can show that the matrix $Q$ is alway primitive. In case $s=0$, 
the resulting theory is just $k$ decoupled  $T[SU(2)]$ theories.

\section{Time-reversal invariant  TQFT}\label{TinvTQFT}

In this section, we will discuss extensively the time-reversal symmetry in TQFTs. We will present several methods to discriminate the time-reversal invariance of Abelian CS theory. All different perspectives turn out to be equivalent, which will be also exemplified in some   time-reversal symmetric  CS theories.

\subsection{Time-reversal invariance from duality}\label{Tinvduality}

In subsection~\ref{CSduality}, we discuss the duality between two different CS theories and propose a set of rules to determine the duality. 

 Now we would like to discuss time-reversal symmetry, which is anit-unitary,  $\sfT: \cA \to \cA$ such that
    \beqn \label{fusionpre}
            \sfT(a\times   b) &=& \sfT(a) \times \sfT(b) ~ ,\qquad \forall a,b \in \cA ~,
         \\  \label{toppre}
            \theta (a) &=& \theta (\sfT(a))^*~,\qquad \forall a \in \cA ~.
            \eeqn
Classically, the time-reversal acts on the CS Lagrangian by flipping the sign of $K$ matrix, namely $\sfT: K \to -K$. 
At quantum level, the CS theory is time-reversal  invariant, or $\sfT$-invariant, if and only if $\mathscr T_{K}$ and $\mathscr T_{-K}$ are dual to each other in the sense we described in section~\ref{CSduality}, namely   the matrix identities  \eqref{dualityconditions} or \eqref{dualitys} are satisfied   for  $K_1=K, K_2=-K$. 

In particular, following the condition  \eqref{dualityconditions3},  there should be integral matrices $R,P$ such that 
\be\label{tinvRP}
R^T K ^{-1} R + K ^{-1}= P~, \qquad P_{ii}\in 2\bZ~,
\ee
which is a necessary condition for time-reversal invariance. 
It is shown in \cite{Delmastro:2019vnj} that a sufficient but not necessary condition for $\sfT$-invariance is that there exits integral matrces $R,P$ satisfying \eqref{tinvRP} and
\be
[K^{-1}RK, R^T]=0~, \qquad [KP,R]=0~.
 \ee
%

 \subsection{Time-reversal invariance from quantum data}
 Following \cite{Belov:2005ze}, we would like to quantize the Abelian CS theory specified by $(\Lambda, K)$, and further discuss the resulting classification of quantum CS theory in terms of the quartet  $(\sD, b,\{\hat q_W\} , \sigma  )$.  In   \cite{Geiko:2022qjy}, the time-reversal invariance is discussed from this perspective. In particular, several methods of discriminating    time-reversal invariance are proposed.
The main goal of this subsection is to review the results in  \cite{Belov:2005ze,Geiko:2022qjy} and   illustrate those methods with explicit examples.  
 
 \subsubsection{Quantization and classification of   Abelian CS theory}

\PA{Classical CS data.} At classical level, the Abelian CS theory is fully specified by the lattice $\Lambda$ and a non-degenerate symmetric bilinear form $K: \Lambda\times \Lambda \to \bZ$. We can choose a basis $e_1, e_2, \cdots$ for the lattice,    then  $\Lambda=\bZ\EV{e_1, e_2, \cdots} \equiv  \{ \sum_i n_i e_i| n_i \in \bZ\}$. The matrix elements of  $K$ are    given by $K_{ij}=K(e_i,e_j)$. We can then write down the classical action for CS theory
\be
S=\frac{1}{4\pi}\int_X K_{ij}A_i dA_j~.
\ee
This formalism based on $K$ matrix is what we used extensively before in this paper (the choice of lattice is implicit in the previous discussions). Therefore the classical CS data is just the pair $(\Lambda, K)$.
In particular, even $K$ gives rise to bosonic CS theory, while odd $K$ gives spin CS theory. More details of CS theory using $K$ matrix  formalism have been discussed in subsection \ref{abeCStheory}.
 
 \PA{Quantum CS data.} The quantum CS data is given by $(\sD, b,\{\hat q_W\} , \sigma  )$.~\footnote{Following \cite{Belov:2005ze}, the data $\sigma$ is more precisely given by  $\sigma \mod 24$. So shifting $\sigma$ by 24 does not change the physical quantum theory, which is consistent with the fact that the modular matrix $T\propto e^{-2\pi i \sigma/24}$ is invariant under such a kind of shift.   } Here $\sD$ is a finite Abelian group, and $b$ is a symmetric non-degenerated bilinear torsion form $b: \sD\times \sD \to \bQ/\bZ$.
The pair $(\sD, b)$ will be referred to as the finite bilinear form. 

We can further define the quadratic refinement for $(\sD, b)$, which is a function $\hat q: \sD \to \bQ/\bZ$ satisfying 
\be\label{qdform}
\hat q(x+y) -\hat q(x) -\hat q(y) +\hat q(0) =b(x,y)~ , \qquad \forall x,y\in \sD~.
\ee  
If a quadratic refinement $q$ further satisfies the homogeneous property $q(nx) =n^2 q(x) $, then $q$ will be referred to as the quadratic form. In particular, this implies $q(-x) =q(x), q(0)=0$. \footnote{We will use hatted $\hat q$ to denote the quadratic refinement, and unhatted $q$ to denote quadratic form. }

Given a quadratic refinement $\hat q$, we can obtain other quadratic refinements. Firstly, we can shift the value $\hat q$ by any rational constant, and define $\hat q'(x) =\hat q(x) +c, c\in \bQ$. Secondly, we can also shift the argument and define $\hat q''(x)=\hat q(x+\delta), \delta \in \sD$. It is easy to check that both $\hat q'$ and $\hat q''$ satisfy \eqref{qdform}, and thus are quadratic refinements for $b$. To avoid the two types of  trivial operations, we can consider the equivalence class of  quadratic refinements subject to the Gauss-Milgram constraint. We say two quadratic refinements   $\hat q_1\sim \hat q_2$ are equivalent if $ \exists \delta \in \sD   $, $\hat q_1(x)=\hat q_2(x+\delta) $ holds for all $\forall x\in \sD$. The resulting equivalence class is denoted as $\{\hat  q_1 \}=\{\hat q_2 \}$. We also need impose the   Gauss-Milgram  constraint
\be\label{MGausssum}
\frac{1}{\sqrt {|\sD|}}\sum_{x\in \sD} e^{2\pi i \hat q(x)}=e^{2\pi i \sigma/8}~,
\ee
for some integer $\sigma\in \bZ$, corresponding to the chiral central charge that we need to specify independently. This constraint \eqref{MGausssum} forbids the freedom of shifting $\hat q\mapsto \hat q+c $ by constant   $  c\in\bQ/\bZ$.

Following  \cite{Belov:2005ze}, the quantum Abelian CS theory is fully specified by the quartet $(\sD, b,\{\hat q \} , \sigma \mod  24)$ subject to the constraints \eqref{qdform} and \eqref{MGausssum}.
The quantum CS theory defined by these data is generally a spin   theory. 
In the special case  that  the quadratic refinement is equivalent to  a quadratic form $q$, then the corresponding CS theory is a bosonic CS theory. Due to the property of quadratic form $q(-x)=q(x), q(0)=0$,  shifting $q$ by   $c$ or $\delta$ would \emph{not}  yield a quadratic from. So there is a unique and preferred quadratic form $q$ in   $\{\hat q \}$. In this case, we can denote the quartet as  $(\sD, b,q, \sigma  ) $ for bosonic CS theory.

\PA{Quantization map.} Now we would like to  understand how to get the quantum CS theory from classical CS theory via quantization.  
Then quantization procedures is just a map from classical CS data $(\Lambda, K)$ to quantum CS data $(\sD, b,\{\hat q_W\} , \sigma  )$. It turns out that the quantization map is surjective, meaning that every quantum CS theory can be obtained from some classical CS theory.

Given the pair $(\Lambda,K)$ for classical CS data, we can first compute    $\sigma$ in  the quantum CS theory, which   is just   the signature   of $K$ matrix,  $\sigma=\text{sig} (K)=\sigma_+-\sigma_-$, where $\sigma _\pm$ are the number of positive/negative eigenvalues of $K$. Note the signature of $K$ is independent of the choice of basis of lattice, and the eigenvalue  of $K$ can never be zero as we have assumed that $K$ is non-degenerate.

Next let us consider the  dual lattice  $\Lambda^*$ defined through the non-degenerate form $K$
\be
\Lambda^*=\Hom(\Lambda, \bZ)= \{x\in \Lambda \otimes \bQ | K(x, \mu) \in \bZ, \forall \mu \in \Lambda\}~.
\ee
Naturally, we also get a dual non-degenerate   bilinear form on the dual lattice $\Lambda^*$
\be
K^*: \Lambda^* \times \Lambda^* \to \bQ~.
\ee
 Since the bilinear form $K$ is integral, we have  $\Lambda \subset\Lambda^*$. Therefore we can  define the discriminant group 
\be
\sD=\Lambda^*/\Lambda~,
\ee
which  is a finite Abelian group  of the form $\prod_i \bZ_{m_i}$, \footnote{After choosing a basis with explicit  $K $ matrix,  the Smith Normal Form of $K$ is just isomorphic to $\sD$, namely $\sD\cong \SNF(K)$. } with order $|\sD| =\prod_i m_i =|\det K|$. 
 The discriminant group  $\sD$ is also  equipped with the torsion form, which is inherited from $K^*$: 
\be
b: \sD \times \sD \to \bQ/\bZ~, \qquad
b(x,y) =K^*(x, y) \mod 1~,
\ee
This is well-defined thanks to the property that    $ b(x+\lambda, y+\mu)  =b(x,y)$ holds for all $\lambda, \mu \in \Lambda$.

We can also determine the quadratic refinement $\hat q$ for $b$ from classical CS data. For this purpose, it is easier to start with the  quadratic refinement $\hat Q: \Lambda \to \bZ$ for $K$, which is defined analogously as  \eqref{qdform}.   One choice of  quadratic refinement of $K$ is given by 
\be\label{QwfromK}
\hat Q_W(\mu) =\frac12 K(\mu, \mu-W) +\frac18 K(W,W) ~,
\ee
 where 
$W\in \Lambda^*$  is  a characteristic vector satisfying 
\be\label{KWvec}
K(W, \mu) =K(\mu, \mu) \mod 2 ~,
\qquad \forall \mu\in \Lambda~.
\ee
 It turns out that up a shift of the argument, the quadratic refinements depend only on $[W]\in \Lambda^*/2\Lambda^*$. For bosonic CS theory with even  $K$, we have a canonical choice $W=0$, and thus $Q(\mu) =K(\mu,\mu)/2$, which is furthermore a quadratic form.

Once we have a quadratic refinement   $\hat Q_W$ of $K$ on $\Lambda$, one can straightforwardly lift it to $\hat Q_W^*$ on the dual lattice $\Lambda^*$, and further  construct $\hat q$ for  $(\sD,b)$. To make the discussion  more explicit, let us pick a basis for $\Lambda=\bZ\EV{e_1, e_2, \cdots}$, then the corresponding symmetric bilinear form is given by a matrix $K_{ij}=K(e_i,e_j)$. The dual lattice can be similarly written as $\Lambda^*=\bZ\EV{e^*_1, e^*_2, \cdots}$ with dual basis $e_i^*$. We can relate the two basis through a matrix $C$:
\be
e^*_i=\sum_j C_{ij}e _j~.
\ee
As a result
\be
K(e^*_i,e_l)=\sum_j C_{ij} K(e_j,e_l)=\sum_j C_{ij} K_{jl}=(CK)_{il}~.
\ee
Therefore, we should choose $C=K^{-1}$, then   $K(e^*_i,e_l)=\delta_{il}$ as expected. This then allows us to relate the two  bases  
\be
e^*_i=\sum_j (K^{-1})_{ij}e _j~, \qquad e _i=\sum_j  K_{ij}e^* _j~.
\ee

Now we need to find the characteristic vector $W$ satisfying \eqref{KWvec}. 
If we write $W=\sum_i W_i e_i^*$ and $\mu=\sum_j \mu_j e_j$, then 
\beqn
K(W, \mu)-K(\mu, \mu)&=&\sum_{ij} W_i \mu_j K(e_i^*, e_j)-\sum_{j,k} \mu_j \mu_k K(e_j,e_k)
\\&=&
\sum_jW_j \mu_j -\sum_{j,k}\mu_j \mu_k K_{jk}
\\&=&
\sum_j W_j  \mu_j -2\sum_{j<k}\mu_j \mu_k K_{jk} -\sum_{j }\mu_j ^2 K_{jj}~,
\eeqn
where we used the fact that $K$ is symmetric.  Further taking the  reduction  mod 2 gives
\be
K(W, \mu)-K(\mu, \mu)  \stackrel{\mod 2\;\;\;\,}{=\!=\!=\!=} \sum_j W_j   \mu_j -\sum_{j }\mu_j  K_{jj}  \stackrel{\mod 2\;\;\;\,}{=\!=\!=\!=}\sum_{j } (W_j-K_{jj})\mu_j     ~,
\ee
where we used the fact that $p^2-p \mod 2=0$ for $p \in \bZ$.  Therefore,  \eqref{KWvec} gives the  condition 
\be\label{KWeven}
W_j-K_{jj}\in 2\bZ~,  \qquad\forall j~.
\ee

Following \eqref{QwfromK},   we can now define the quadratic refinement of $K$ explicitly for any   $\mu\in \Lambda$, 
\be
\hat Q_W(\mu) =\frac12 K(\mu, \mu-W)  +\frac18K(W,W)
=\frac12 \bm\mu^T K\bm\mu - \frac12\bm \mu^T\bm W +\frac18\bm W^T K^{-1}\bm W~,
\ee
where $\bm\mu,\bm W$ are regarded as column vectors with entries $\mu_i$ and $W_i$, and $K,K^{-1}$ between vectors should be understood as    matrices. 

For  any element in the dual lattice $x\in \Lambda^*$, we can write it as $x=\sum_i x_i e_i^*=\sum_{ij}  (K^{-1}\bm x)_{ij}e_j $, then  the quadratic refinement on the dual lattice is
\be
\hat Q^*_W(x)
=\hat Q_W(\sum_{i}( K^{-1} \bm x)_ie_i)
=\frac12\bm x^T K^{-1}\bm x - \frac12\bm  x^T K^{-1}\bm W +\frac18\bm  W^T K^{-1}\bm W~.
\ee
Consequently, the quadratic refinement $\hat q$ on $\sD$ is given by 
\be\label{qWx}
\hat q _W(x) =\frac12\bm x^T K^{-1}\bm x - \frac12\bm  x^T K^{-1}\bm W +\frac18\bm  W^T K^{-1}\bm W \mod 1~, \qquad
x\in \sD~.
\ee
which is indeed well defined becuase it is invariant if under the   shift   $\bm x \mapsto \bm x+ K \bm y$ for any $\bm  x, \bm y\in \bZ^n$  thanks to the condition \eqref{KWeven}.
The bilinear form on $\sD$ then follows straightforwardly
\be\label{bilform}
b(x,y) =\hat q _W(x+y)+\hat q _W(0)-\hat q _W(x)-\hat q _W(y)=\bm x^T K^{-1}\bm y~,
\ee
which is independent of $W$.  It is also possible to  show that \eqref{MGausssum} is satisfied for the quadratic refinement \eqref{qWx}. 

Given a   $K$ matrix,  one can extend $K$ by one diagonal $ \pm 1$, and get a new $K$ matrix $ \PBK{ K & 0 \\ 0 & \pm 1}$.  In the quantization, we also need to extend the characteristic vector $\bm W$ by adding one more odd entry     for the consistency of  $W_j-K_{jj}\in 2\bZ$.  However, this does not modify the group $\sD$. Consequently,  the only effect is that  $\sigma$ is shifted by $\pm 1$, and $\hat q _W(x) $ is shifted by $(\pm 1 \mod \bZ)/8$, consistent with \eqref{MGausssum}. Therefore, the  extension of $K$ matrix by  $ \pm 1$ looks a bit trivial. 
However, physically, there is a big difference if the original $K$ matrix is even, corresponding to bosonic TQFT. In that case, we have a canonical choice of quadratic refinement, which is actually a quadratic form. After   extending $K$ by one diagonal $ \pm 1$, then  resulting theory is spin, and it is impossible to assign a quadratic form.

 \subsubsection{Classification of time-reversal invariant Abelian  CS theory}\label{tinvgeneral}
Following previous discussions, we learn that  the classification of Abelian CS theory reduces to the classification of the quartet $(\sD, b,\{\hat q_W\} , \sigma  )$. Since any  finite Abelian group $\sD$ is isomorphic to 
$
 \bZ_{p_1^{d_1}}\times \cdots \times  \bZ_{p_t^{d_t}}
$
 where each $p_j$ is prime but not necessarily distinct, we can study each factor $\bZ_{p^d}$ independently. 
So the  problem reduces to the classification of bilinear form and its quadratic refinement/form on $\sD$. In particular, the  generators of quadratic form for each  factor $\bZ_{p^d}$ have been classified, which is given in table \ref{quadform}.
\begin{table}[h]
\[\def\arraystretch{1.2} 
\begin{array}{|c|c|c|   }  
    \hline
    \text{Symbol} & \text{Abelian  group} & \text{Quadratic form}  \\
    \hline
    \lambda_{p^r}^{+1} &  \mathbb{Z}_ {p^r}, \; r \geq 1 & q(x) = ux^2 / p^{r+1} \\
    \lambda_{p^r}^{-1} &  \mathbb{Z}_{p^r}, \; r \geq 1 & q(x) = vx^2 / p^{r+1} \\
      \omega_{2^r}^{\pm 1} &  \mathbb{Z}_{2^r}, \; r \geq 1 & q(x) = \pm x^2 / 2^{r+1} \\
    \omega_{2^r}^{\pm 5} & \mathbb{Z}_{2^r}, \; r \geq 2 & q(x) = \pm 5x^2 / 2^{r+1} \\
    \phi_{2^r} & \;  \mathbb{Z}_{2^r} \times \mathbb{Z}_{2^r}, \; r \geq 1 \qquad& q(x,y) = xy / 2^r \\
    \psi_{2^r} & \; \mathbb{Z}_{2^r} \times \mathbb{Z}_{2^r},\; r \geq 1 \qquad  &\quad q(x,y) = (x^2 + xy + y^2) / 2^r \quad\\
    \hline 
\end{array} 
\] 
\caption{Generators of quadratic form. Here $p$ is an odd prime, and  $2u $  $(2v)$ is a quadratic residue (non-residue) mod $p$.}\label{quadform}
\end{table}

To study time-reversal invariant Abelian CS theory, we need to impose further constraints.   Since $\sfT: (\sD, b,\{\hat q_W\} , \sigma  \mod 24  ) \mapsto (\sD,- b, \{-\hat q_W\} , -\sigma  )$, we get    time-reversal invariance if the two quartets are actually isomorphic.  We will refer to such a kind  of  quartet as $\sfT$-symmetric quartet. We can similarly define bilinear $\sfT$-form, quadratic $\sfT$-refinement, and  quadratic $\sfT$-form. We will not give their precise definitions here, but refer the readers to \cite{Geiko:2022qjy} for details.
It turns out that every bilinear $\sfT$-form $b_\sfT$ together with $\sigma=0,4 $ admits an equivalent class of quadratic $\sfT$-refinement $\{\hat q_\sfT\}$; they can be furthermore explicilty represented   either  as $\hat q_\sfT=\omega_2^{+1}-1/8=x^2/4-1/8$, and $\hat q_\sfT=\omega_2^{+1}+3/8=x^2/4+3/8=x^2/4-1/8+1/2$ in the special case $\sD=\bZ_2$,  or as $\hat q_\sfT=q_\sfT$ and $\hat q_\sfT=q_\sfT+1/2$ in other cases. \footnote{This implies that all the $\sfT$-invariant spin TQFT is given by $\mathscr T\boxtimes  \big(U(1)_2  \big)^{\boxtimes r}\boxtimes \sVec$ where $\mathscr T$ is some $\sfT$-invariant bosonic TQFT. Note $\omega_2^{+1}$ is just the quadratic refinement for $U(1)_2   $ CS theory, which is not $\sfT$-invariant as bosonic TQFT, but $\sfT$-invariant as spin TQFT (or more precisely, $U(1)_2 \boxtimes \sVec$ is $\sfT$-invariant.). Furthermore, shifting $\hat q_\sfT$ by $1/2$ would change the chiral central charge by $4$, which does not spoil the time-reversal invariance after stacking with $E_8$ model.} Here $q_\sfT $ is a quadratic $\sfT$-form, satisfying $(\sD, q_\sfT) \cong( \sD, -q_\sfT)$.  Given the generators in table  \ref{quadform}, one can then verify the $\sfT$-invariance and construct  quadratic $\sfT$-forms explicitly.  It turns out that the  $\sfT$-invariance can be checked easily by computing the Gauss sum \cite{Geiko:2022qjy}.

 \subsection{Time-reversal invariance from reality of Gauss sums and generating function}

As proved in  \cite{Geiko:2022qjy}, the  $  (\sD, b,\{\hat q_W\} , \sigma     )$ is $\sfT$-symmetric    if and only if all the    Gauss sums are real.  Here the  $n$-th Gauss sum is defined as
\footnote{The summation is over half of the anyons in the spin case.  }
\be
\tau_n(\sD,\hat q)=\frac{1}{ \sqrt{|\sD|}}\sum_{x\in \sD}e^{2\pi i n \hat q (x)}~, \qquad
\tau_1(\sD,\hat q)=e^{2\pi i \sigma/8}~.
\ee
This   gives the necessary and sufficient condition to check the  time-reversal invariance of abelian CS theory  theory.~\footnote{From the reality of  $\tau_1$, we learn that $\sigma\equiv 0,4 \mod 8$. This seems to be different from the condition $\sigma \equiv -\sigma \mod 24$, which implies  $\sigma\equiv 0,12 \mod 8$. However, this   discrepancy can be simply and unambiguously  remedied by stacking with Kitaev's $E_8$ model if necessary.  }
For bosonic CS theory, the quadratic refinement $ \hat q$ can be chosen to be a quadratic form $q$.

For computational purpose, it is helpful to rewrite the formula in terms of the $K$ matrix using \eqref{qWx}
\be\label{taunKw}
\tau_n(K )=\frac{1}{\sqrt{|\det K|}}\sum_{\bm\alpha\in \bZ^k/K \bZ^k }
e^{2 \pi i n\hat q_w( \bm \alpha) }~,
\ee
where
\be\label{qwalpha}
\hat q_w( \bm \alpha)=\frac12\bm \alpha^T K^{-1} \bm \alpha -\frac12\bm \alpha^T K^{-1} \bm w +\frac18\bm w^T K^{-1}\bm w
 =\frac12(\bm \alpha-\frac12\bm w)^T K^{-1}(\bm \alpha-\frac12\bm w) ~.
\ee 
Here $\bm w\in \bZ^k $ is a characteristic vector satisfying  $\bm \gamma^T K  \bm \gamma-\bm \gamma^T \bm w\in 2 \bZ$  for all $\bm \gamma\in \bZ^k$.  More explicitly, 

\be
\bm \gamma^T K  \bm \gamma-\bm \gamma^T \bm w
=2\sum_{i<j}\gamma_{i}\gamma_j K_{ij}+\sum_i \gamma_i^2 K_{ii}-\sum_i \gamma_i w_i
 \stackrel{\mod 2\;\;}{=\!=\!=\!=} \sum_i \gamma_i ( K_{ii}  - w_i)~.
\ee
So the actual condition is \be
w_i -K_{ii}\in  2\bZ~,
\ee 
 and for simplicity we can just take $w_i =K_{ii}$.  
For even $K$, the canonical choice is $\bm w=0$ and the summand becomes the power of topological spin $e^{2 \pi i n\hat q_w( \bm \alpha) }= \theta({\bm \alpha})^n$.   Note the Gauss sum is independent of the  specific choice of $\bm w$  as long as $w_i -K_{ii}\in  2\bZ$ due to  the summation   over all $\bm\alpha\in \bZ^k/K \bZ^k$ and   the rewriting in  \eqref{qwalpha}. 
Under the shift $\bm \alpha \mapsto  \bm \alpha+K \bm \gamma$, it is easy to see that
\be
\hat q_w( \bm \alpha+K\bm\gamma)-\hat q_w( \bm \alpha)=\frac12\Big(
  \bm \alpha^T \bm \gamma+  \bm \gamma^T \bm \alpha  
 +\bm \gamma^T K \bm \gamma-\bm\gamma^T \bm w \Big)
\in  \bZ~.
\ee
Therefore the Gauss sum defined in \eqref{taunKw} by  summing over $\bZ^k /K \bZ^k$  is unambiguous.  


 Therefore the  CS theory defined through $K$ matrix    is time-reversal invariant  after quantization if and only if  $\tau_n(K )\in \bR$ for all $n \in \bZ$.

However, it may happen that a time-reversal non-invariant  CS theory becomes time-reversal invariant after we stack with some copies of \sVec. The procedure of stacking $\sVec$ is equivalent to extending $K$ matrix by some diagonal entries $\pm 1$. It would promote a bosonic CS theory to a spin CS theory and furthermore change the central charge by some integer.  After quantization, this would change the quantum CS data by shifting  the quadratic refinement by some constant. We can trivialize such a kind of constant shift in $\hat q$ by considering the reduced Gauss sums
\be
\eta _n(\sD, q)=
\frac{\tau_n(\sD, q)}{\big  (\tau_1(\sD,q) \big)^n} ~, \qquad \text{ or }\qquad
\eta _n(K )=
\frac{\tau_n(K, \bm w)}{\big  (\tau_1(K, \bm w) \big)^n} ~.
\ee 
Then  a CS theory is time-reversal invariant after stacking with some copies of $\sVec$  if and only if  $\eta _n(\sD, q)$ or $\eta_n(K )\in \bR$ for all $n \in \bZ$.

It is simpler to introduce a generating function for the Gauss sum, which will be referred to as Gauss generating function 
\beqn
\Upsilon(\sfz)&=&
\sum_{j=0}^\infty\tau_ {j } \sfz^ {j } =\frac{1}{ \sqrt{|\sD|}}\sum_{j=0}^\infty \sfz^ {j }  \sum_{x\in \sD   }e^{2 \pi i j \hat q (x)}
=\frac{1}{ \sqrt{|\sD|}} \sum_{x\in \sD   } \sum_{j=0}^\infty \Big( \sfz    e^{2 \pi i   \hat q (x)} \Big)^j
\\&=&
\frac{1}{ \sqrt{|\sD|}} \sum_{x\in \sD   }  \frac {1}{  1- e^{  2\pi i   \hat q (x)}\sfz }~.
\label{GausGene}
\eeqn
The reality of the Gauss  generating function is equivalent to the reality of all the Gauss sums; both are further equivalent to the time-reversal invariance of CS theory.  One can similarly define the reduced Gauss generating function for $\eta$. 
%

Stacking   two TQFTs $K$ and $ K'$ gives another TQFT with level matrix $K\oplus K'$, whose   Gauss  sum is given by the product 
\be
\tau_n({K\oplus K'})=
\tau_n(K)   \tau_n(K')~,
\ee
and the Gauss
generating function is given by the     Hadmard product
 \be
  \Upsilon_{ {K\oplus K'} }=
 \Upsilon_K *  \Upsilon_{K'}~.
 \ee

 \subsection{Time-reversal invariance from self-perpendicularity of lattice  }\label{selfperp}
 It is further shown in \cite{Geiko:2022qjy} that the CS theory $(\Lambda, K)$ is time-reversal invariant after quantization if and only if the lattice is self-perpendicular $(\Lambda, K) \perp (\Lambda, K)$.
 
Two lattices  are perpendicular $\Lambda\perp \Lambda'$  if there exists a unimodular lattice $L$ with bilinear form $g$ and a primitive embedding  \footnote{This means $L/\Lambda$ is also a lattice. The primitive embedding is equivalent to the primitivity of embedding matrix. As a non-example,  the embedding of $2\bZ$ into $\bZ$ is not    primitive  because $\bZ/2\bZ=\bZ_2$ which is torsional and thus not a lattice.}  of $\Lambda$ into $L$ such that $\Lambda'\cong (\Lambda)_{L}^\perp  $, which is the orthogonal complement of $\Lambda$ in $L$ with respect to $g$. 

To be more explicit, we can choose a basis for the lattices: $L=\bZ\EV{v_1, \cdots}$,
$\Lambda=\bZ\EV{e_1, \cdots}$, $\Lambda'=\bZ\EV{e_1', \cdots}$. 
Since $\Lambda $ embeds into $L$, we have  
$e_I=\sum_i E_{Ii}v_j$, $K_{IJ}=\sum_{ij}E_{Ii}E_{Jj}g_{ij}$, where $g_{ij}=g(v_i,v_j)$. Similarly, we also have
$e'_I=\sum E'_{Ii}v_i$, $K'_{IJ} =\sum_{ij} E'_{Ii}E'_{Jj}g_{ij}$. 
Since $L$ is unimodular,   we have the condition $|\det g|=  1$. Primitive embedding means the matrices $E,E'$ are primitive in the sense described in subsection~\ref{chargematrx}. Then we get some matrix identities
\be
K=E gE^T , \qquad   K'=E' gE'^T, \qquad Eg E'^T=E g E^T=0~, \qquad\det\cE \neq 0~,
\ee
where $\cE\equiv   \PBK { E \\  E' } $ is a square matrix constructed out of the two embedding matrices, and 
the last  two equalities follows from  the orthogonality and complementarity  of two lattices $\Lambda'\cong (\Lambda)_L^\perp  $.

In the  self-perpendicular case,  the two lattices $\Lambda$ and $\Lambda'$ should be isomorphic. In particular, we are supposed to be able to choose a   basis for $e_I'$ via a proper unimodular transformation such that $K' =K $.
Therefore the self-perpendicular condition of  $(\Lambda, K)$ are represented as
\beqn\label{ABK}
K=E gE^T =E' gE'^T~, \qquad E' g E^T=E g E'^T=0~, \qquad |\det g|=1~, \\ \qquad 
E,E' \text{ are primitive matrices}~, \qquad \det \cE\equiv\det  \PBK { E \\  E' }  \neq 0~ .
\eeqn
 Actually the last non-degenerate condition can be derived from the rest of conditions. Indeed, 
 \be
 \cE g \cE^T= \PBK { E \\  E' }  g \PBK { E^T \;  E'^T } = \PBK { Eg E^T \; Eg E'^T\\  E'gT^T \; E' g E'^T } 
 = \PBK { K \; 0 \\  0 \; K } 
 \ee
 Therefore $(\det \cE)^2 \det g =    (\det K)^2$. This implies $\det \cE \neq 0$ as long as $K$ is non-degenerate which is alway assumed.  It also implies that $\det  g=1$.

\subsection{Examples }\label{CSTinvexample}
In this subsection, we will give some examples of time-reversal invariant CS theory. In particular, we will try to use these examples to exemplify   different  criterions of time-reversal invariance we established before. 
 
\PA{$U(1)_k$ CS theory. } Let us start with the simplest example, $U(1)_k$ CS theory. More precisely, we consider CS theory with $K$ matrix   $K=\PBK{k & 0 \\ 0 & -1}$, $k>0$, where the stacking with    $U(1)_{-1}$  guarantees the chiral central charge  vanishes $\sigma=0$.  As shown in \cite{Delmastro:2019vnj}, this theory   is time-reversal invariant  if and only if   $k \in \bT$ where
  \beqn
 \mathbb T &=&\{k \in \bZ_+  \; |\; kp-q^2=1 \text{ for some } p,q\in \bZ \}
\\&=& \label{tvalue}
\{ k\in \bZ_+   \; |\; \exists \text{   coprime integers } a,b\in \bZ,\gcd(a,b)=1 \text{ such that }k=a^2+b^2 \}
\\&=&
\mathbb T_\text{odd}\cup 2 \mathbb T_\text{odd}
\\&=&
\{  1, 2, 5, 10, 13, 17, 25, 26, 29, 34, 37, 41, 50, 53, 58, 61,65, 73, 74, 82, 85, 89, 97, \cdots\}~.
 \eeqn
 For later convenience, we also introduce  the  set
 \beqn\label{Todd}
\mathbb T_\text{odd} &=&\{k \in \bZ_+  \; | k=\prod_{p_i:\text{ prime and } p_i =1 \mod 4  }p_i^{r_i} \}
=
\{  1,   5,   13, 17, 25, 29,   37, 41,   53,  61,65, 
 \cdots\} ~.
 \qquad\qquad
\eeqn
 The discriminant  group is $\sD=\bZ^2/K \bZ^2 \cong\bZ_k$. Then we can construct the quadratic form and bilinear form according to \eqref{qWx} and \eqref{bilform}
 \be\label{hatq1}
  \hat q'(x)= \frac{x^2-xk}{2 k} +\frac{k-1}{8}   = \frac{(x -k/2)^2}{2 k} -\frac{  1}{8} ~  , \qquad b'(x,y)=\frac{xy}{k}~.
 \ee 
 
 Alternatively,    the quadratic refinement and bilinear form  can be  chosen as
   \footnote{The quadratic refinements are given in   \cite{Geiko:2022qjy}, up to a missing factor  of  2.}
 \be\label{hatq2}
 \hat q(x) =\begin{cases}
 \frac{x^2}{2k}-\frac18~,\quad &   k \text{   even }\\
  \frac{2x^2}{ k}+\frac k 8-\frac18~,\quad & k \text{   odd }\\
 \end{cases}~,
\qquad
b(x,y) =\begin{cases}
 \frac{xy}{ k}~,\quad  & k\text{   even }\\
  \frac{4xy}{ k} ~,\quad & k \text{   odd }\\
 \end{cases}~,
\qquad
x,y \in \bZ_k~.
 \ee
 Let us show that \eqref{hatq1} and \eqref{hatq2} are indeed equivalent.   
 For even $k$, obviously we have  $ \hat q'(x+k/2)=\hat q(x)$, so $\hat q'$ and $\hat q$ are two equivalent quadratic refinements for   the same $b$. 
 For odd $k$, we can consider the group homomorphism $\bZ_k\to \bZ_k$ defined by $ x\mapsto 2x$. Then we   have $b'(2x,2y)=4xy/k$ and 
 \be
 \hat q'(2x)= \frac{4x^2-2xk}{2 k} +\frac{k-1}{8} \stackrel{\mod 1\;\;\;\,}{=\!=\!=\!=} \frac{2x^2 }{  k} +\frac{k-1}{8}~,
 \ee
which  is indeed equivalent to \eqref{hatq2}.
 
One can also   check that \eqref{hatq2} are  also quadratic $\sfT$-refinements. In particular, the constant shifts are either $-1/8$ \footnote{For even $k $, there is precisely one $\bZ_2$ factor in the discriminant group $\sD=\bZ_2\times \bZ_{k/2}$ as $\gcd(k/2,2)=1$ for $k\in \bT$.  Therefore, it must involve the quadratic $\sfT$-refinement $\omega_2^{+1}-1/8$  or $\omega_2^{+1}+3/8$, which is the origin of $-1/8$ shift in eq.~\eqref{hatq2}. } or $(k-1)/8\equiv 0,1/2 \mod 1$ for $k \in \bT_{\text{odd}}$, which are consistent with the general result in subsection~\ref{tinvgeneral}. One can also verify that the quadratic parts in \eqref{hatq2}  are indeed quadratic $\sfT$-form. 
 

 Now we switch to the lattice perspective.  We start with the  lattice $L=\bZ\EV{v_1, v_2, v_3, v_4} $  with bilinear form $g=\diag(1,1,-1-1)$.  So $L$ is unimodular with $\det g=1$.     We can  consider two lattices $\Lambda=\bZ\EV{e_1, e_2},\Lambda'=\bZ\EV{e'_1, e'_2}$, and embed them  into $L$ as follows:
 \beqn
 &&e_1=m v_1+n v_2~, \qquad e_2=v_3~,  \qquad  E=\PBK{m & n & 0 &0 \\ 0 & 0 &1 & 0}~,
\\  &&
 e_1'=m' v_1+n' v_2~, \qquad e_2'=v_4~, \qquad  E=\PBK{m' & n' & 0 &0 \\ 0 & 0 &0 & 1}~.
 \eeqn
 It is then easy to compute that 
 \beqn
 K&=&EgE^T=\PBK{m^2+n^2 & 0 \\ 0 & -1}~, \qquad K'=E'gE'^T= \PBK{m'^2+n'^2 & 0 \\ 0 & -1}~,
\\   
S&=&E'gE^T= E gE'^T=  \PBK{mm' +nn'  & 0 \\ 0 & 0}~.
 \eeqn
 In order to get self-perpendicular lattice, we should impose the primitive embedding condition, together with $K=K'$ and $S=0$. Therefore $m^2+n^2=m'^2+n'^2$ and $mm'+n n'=0$ which can be solved by $m'=sn, n'=- sm$ subject to $s^2 =1$. The primitive embedding condition yields $\gcd(m,n)=\gcd(m',n')=1$. 
  This reproduces exactly the set of time-reversal invariant theory with  $k\in \bT$ we discussed before. 
 
    By direct computation, one  can verify that    the Gauss sums of the  $U(1)_k$ spin CS theory     are  real if $k \in \bT$.  For illustration, we give the explicit  Gauss generating functions \eqref{GausGene}    for several different values of $k$:  
      \beqn
k=2~, \qquad && \Upsilon=\frac{\sqrt{2}-\sfz}{ 1-\sqrt{2} \sfz+\sfz^2}~,
\\
k=3~, \qquad && \Upsilon=\frac{-3 i \sfz-\sqrt{3} \sfz+6}{(\sfz+i) \left(3 i \sfz+\sqrt{3} \sfz-2 i \sqrt{3}\right)}~,
\\
k=5~, \qquad && \Upsilon= \frac{\sfz \left(-2 \sfz+\sqrt{5}-3\right)+2 \sqrt{5}}{(\sfz+1) \left(2 \sfz^2-\left(\sqrt{5}+1\right) \sfz+2\right)} ~.
\label{Gsk5}
  \eeqn
%
So we see that the  Gauss generating function is indeed real for $k=2,5\in \bT$, and complex for $k=3\not\in \bT$.

\PA{$SU(5)_1$ CS.} Let us now consider the $SU(5)_1$ CS, which is equivalent to the  $U(1)$ CS theory with $K$ matrix given by the Cartan matrix of group $A_4\simeq SU(5)$. The corresponding $A_4$ lattice can be obtained as the sub-lattice of the   $E_8$ lattice. The  $E_8$ lattice is unimodular and admits the bilinear form specified by  
\be
g= 
\left(
\begin{array}{cccccccc}
 4 & -2 & 0 & 0 & 0 & 0 & 0 & 1 \\
 -2 & 2 & -1 & 0 & 0 & 0 & 0 & 0 \\
 0 & -1 & 2 & -1 & 0 & 0 & 0 & 0 \\
 0 & 0 & -1 & 2 & -1 & 0 & 0 & 0 \\
 0 & 0 & 0 & -1 & 2 & -1 & 0 & 0 \\
 0 & 0 & 0 & 0 & -1 & 2 & -1 & 0 \\
 0 & 0 & 0 & 0 & 0 & -1 & 2 & 0 \\
 1 & 0 & 0 & 0 & 0 & 0 & 0 & 2 \\
\end{array}
\right)~,
\qquad
\det g=1~.
\ee
Now we can consider two sub-lattices $\Lambda, \Lambda'$ specified by the  following  primitive embedding matrices
\be
E=\left(
\begin{array}{cccccccc}
 0 & 0 & 0 & 1 & 0 & 0 & 0 & 0 \\
 0 & 0 & 0 & 0 & 1 & 0 & 0 & 0 \\
 0 & 0 & 0 & 0 & 0 & 1 & 0 & 0 \\
 0 & 0 & 0 & 0 & 0 & 0 & 1 & 0 \\
\end{array}
\right), 
\qquad
E'=\left(
\begin{array}{cccccccc}
 -3 & -6 & -5 & -4 & -3 & -2 & -1 & 1 \\
 3 & 6 & 5 & 4 & 3 & 2 & 1 & -2 \\
 1 & 1 & 0 & 0 & 0 & 0 & 0 & 0 \\
 -4 & -6 & -5 & -4 & -3 & -2 & -1 & 2 \\
\end{array}
\right) ~,
\ee
and   compute
\be 
K=EgE^T=E'g E'^T =C_{A_4}=\left(
\begin{array}{cccc}
 2 & -1 & 0 & 0 \\
 -1 & 2 & -1 & 0 \\
 0 & -1 & 2 & -1 \\
 0 & 0 & -1 & 2 \\
\end{array}
\right)~,
 \qquad\qquad 
 Eg E'^T=E g E^T=0~.
\ee

This construction tells us that the U(1) CS theory with $K$ matrix above is time-reversal invariant as bosonic theory: the $K$ matrix has even diagonals  only and there is no need to stack with any transparent fermion.

From the $K$ matrix above, it is easy to see  $\sigma=4$ and $\sD=\bZ_5$ with quadratic form $q(x) =2x^2/5$ and bilienar form $b(x,y)=4xy/5$.  One can check that  all the Gauss sums are   real. In particular, we can compute the Gauss generating function 
\be
\Upsilon =\frac{\sfz \left(2 \sfz+\sqrt{5}-3\right)-2 \sqrt{5}}{(\sfz-1) \left(\sfz \left(2 \sfz+\sqrt{5}+1\right)+2\right)}~.
\ee
 This is  the same as \eqref{Gsk5} up to   the sign flip of $\sfz$, and can be regarded as the consequence of level-rank duality   between   $SU(5)_1$  and  $U(1)_{-5}$   CS  theory as   spin TQFTs.  
 
\PA{Minimal TQFT $\cA^{N,p}$. } Another simple class of   abelian TQFT is given by the so-called minimal TQFT \cite{Hsin:2018vcg}, denoted as  $\cA^{N,p}$ where the two positive integers are subject to the coprime condition $\gcd(N,p)=1$. The anyons of this  theory  are generated by  $W$, and the general anyon can be denoted as $W^k$ satisfying abelian fusion rules.  The topological spin is 
\be
\theta_k=\theta(W^k)=e^{2\pi i pk^2/2/N }~.
\ee
When $pN$ is even,  
we get a bosonic theory and $W^N=\bf 1$.  \footnote{When $pN$ is odd, we get a spin theory.} 
The condition  $\gcd(N,p)=1$ guarantees that the corresponding $S$-matrix is non-degenerate.  

  As shown in \cite{Delmastro:2019vnj}, the minimal TQFT $\cA^{N,p}$ is time-reversal invariant if and only if  
\be
p\in\mu(N) \bZ~,
\ee
where
\be \label{muN}
\mu(N)=N\Big/\prod_{i : \text{ prime }p_i =1 \mod 4} p_i^{r_i}~, \qquad\qquad N=\prod_{i: \text{ prime } p_i} p_i^{r_i}~.
\ee
Actually time-reversal invariance can only be possible when $\mu(N)=1$, because otherwise 
  $\gcd(p,N)>1$ for $p=m \mu(N)  $ and $ \mu(N) >1$. It is easy to see that $\mu(N)=1$ is equivalent to $N\in \bT_{\text{odd}}$ \eqref{Todd}. 
$\gcd(p,N)=1$ further implies that $\gcd(p,p_i)=1$ for any prime factor   $p_i$ of $N$. As a result, $\cA^{N,p}$ is time-reversal invariant if and only if  
\be\label{ApNTinv}
N=\prod_{p_i:\text{ prime, } p_i =1 \mod 4, \text{ }\gcd(p_i, p)=1  }p_i^{r_i}  
 =\bT_\text{odd}\backslash\Big(\bigcup_{q|p} q\bZ \Big)~.
\ee
Let us show that the same condition can be obtained from the reality of Gauss sums in the bosonic  $\cA^{N,p}$ TQFTs. Since $Np\in 2\bZ$ and $\gcd(p,N)=1$, there are two possibilities: $p$ even, $N$ odd, or the opposite. 

For odd $N$    and  even $p=2q$  subject to  $\gcd(N,q)=1$,   the Gauss sums are 
\footnote{The quadratic Gauss sum function is defined as
\be
 G(a,d) 
 =\sum_{j=0}^{d-1}e^{2\pi i a j^2/d}~,
 \qquad
  G(a,d) =\gcd(a,d)\; G(\frac{a}{\gcd(a,d)},\frac{d}{\gcd(a,d)})
 \ee 
 In the coprime case $\gcd(a,d)=1$, the  quadratic Gauss sum  is  explicitly given   by  
 \begin{equation*}
G(a,d) =  
\begin{cases} 
0 & \text{if } d \equiv 2 \pmod{4} \\
\varepsilon_d \sqrt{d} \left( \frac{a}{d} \right) & \text{if } d \equiv 1 \pmod{2} \\
(1+i) \varepsilon_a^{-1} \sqrt{d} \left( \frac{d}{a} \right) & \text{if } d \equiv 0 \pmod{4}
\end{cases}
\qquad~,\qquad
\varepsilon_m =  
\begin{cases} 
1 & \text{if } m \equiv 1 \pmod{4} \\
i & \text{if } m \equiv  3 \pmod{4} \\
 \end{cases}~,
\end{equation*}
 where $ \left( \frac{a}{d} \right)$ is Jacobi symbol.  Note $\left( \frac{a}{d} \right)=\pm 1$ for $\gcd(a,d)=1$. For more details, see  \cite{gaussum}.
 \label{Gsum}
 }
\be
\sqrt{N}\tau_m=
\sum_{k=0}^{N-1}\theta_k^m=\sum_{k=0}^{N-1} 
e^{2\pi im qk^2/ N }
=G (qm,  N)=l G(qm/l,  N/l)
=l \varepsilon_{N/l}\sqrt{N/l} \left( \frac{qm/l}{N/l} \right) ~,
\ee
  where $l =\gcd(qm,  N)=\gcd( m,  N)$, and  $\gcd(qm/l,  N/l)=1$, which implies $ \left( \frac{qm/l}{N/l} \right) =\pm 1$.  Then the reality   of the Gauss sums  is translated to the condition that    $\varepsilon_{N/l}$ is real for all $l$, which means 
\be
N/l=N/\gcd( m,  N)=1 \mod 4~, \qquad\forall  m \in \bN~.
\ee
It is easy to convince oneself that  this is equivalent to the condition   $ \mu(N)=1$ where $\mu(N)$ is defined in \eqref{muN}.

For even $N  $, and odd $p $,   the Gauss sums are 
\be
2\sqrt{N}\tau_m=
2\sum_{k=0}^{N-1}\theta_k^m=
\sum_{k=0}^{2N-1}\theta_k^m=\sum_{k=0}^{2N-1} 
e^{2\pi im pk^2/ 2/N }
=G (pm, 2 N)=l G(pm/l,  2N/l)~,
 \ee
 where we used the fact that $e^{2\pi im pk^2/ 2/N }=e^{2\pi im  (p+N)k^2/ 2/N }$ for even $N$, and
$  l =\gcd(pm, 2 N)=\gcd( m, 2 N)$. For $m=1$, we thus have $l=1$ and   $2\sqrt N \tau_1=G(p,2N)$. Since  $2N\equiv 0 \mod 4$, $G(p,2N)\propto (1+i)$ is always complex (see footnote \ref{Gsum}).   This implies that   the reality condition of  Gauss sum are never satisfied    in this case with  even $N$. In particular,  $\mu(N)$ is also  even. 

Combing the two possibilities above, we get the condition $\mu(N)=1$ and $\gcd(p,N)=1$ for time-reversal invariance, in agreement with \eqref{ApNTinv}.

\section{Time-reversal invariance from self-mirror symmetry} \label{SCFTtoTQFT}
This section gives the main result of this paper, and is devoted to establishing the connections between  self-mirror symmetry in SCFT and time-reversal symmetry  in TQFT. 
The relation can be simply represented as follows:
\be \boxed{ 
\text{Self-mirror symmetric   SCFT } \cT_{Q}\quad
\xrightarrow {\text{universal mass def.}}\quad
\text{Time-reversal invariant   TQFT }\mathscr T_{K=QQ^T}
}\label{selfduals}
\ee
It can be regarded as a special case of figure \ref{mirrorscfttqft}, where the left and right columns are actually equivalent in the senses   that we have discussed before.

\subsection{General proof}
\subsubsection{$\sfT$-invariance from the duality perspective   }
As we discussed in section \ref{mirrorduality}, the mirror symmetry in SCFTs gives rise to the duality in TQFTs after performing the universal mass deformation. 
More precisely, the mirror pair of SCFTs $\cT_Q, \cT_{\wt Q}$ satisfies $\wt Q Q^T=0$, and yields  the duality  pair of  TQFTs $\sT_K, \sT_{\wt K}$ with  $K=QQ^T, \wt K =-\wt Q \wt Q^T$. Here the sign in $\wt K$ comes from the sign flipping of mass parameter in mirror symmetry.

For self-mirror symmetric SCFTs  $\cT_Q$, the charge matrix should satisfy \eqref{selfMirror}
$Q \Omega Q^T=0$ for some signed permutation matrix $\Omega$.
Therefore, we can take $\wt Q =Q \Omega$, then 
\be
-\wt K=\wt Q \wt Q^T=Q \Omega \Omega^T Q^T=QQ^T=K~.
\ee
Since by our construction, the two TQFTs $\sT_K, \sT_{\wt K  }$  are dual to each other, we conclude that $\sT_K$ is dual to $\sT_{-K}$, namely $\sT_K$ is time-reversal invariant, following the discussion in subsection~\ref{Tinvduality}. 

Therefore, as a special case of duality, we have proved that time-reversal invariance follows from self-mirror symmetry under universal mass deformation.
 \subsubsection{$\sfT$-invariance from the lattice  perspective }
 The connection between time-reversal invariance and self-mirror symmetry can also be proved from the lattice perspective. 
As we discussed in subsection~\ref{selfperp}, time-reversal invariance is equivalent to the self-perpendicularity of the lattice $(\Lambda, K)$.  This further boils down to   the conditions in \eqref{ABK}
 \be\label{lattembed}
K=E gE^T =E' gE'^T~, \qquad E' g E^T=E g E'^T=0~, \qquad |\det g|=1~ ,
 \ee
for primitive   matrices $E,E' $ which describes the embedding of  $\Lambda$ into an unimodular lattice with bilinear form $g$.

In the current case of interest, we can choose
\be
E=Q~, \qquad E' =Q \Omega~, \qquad g=\mathds{1}~,
\ee
then \eqref{lattembed} yields 
\be
K=QQ^T=Q\Omega\Omega^TQ^T, \qquad Q\Omega^T Q^T=Q\Omega  Q^T=0~.
\ee
This is exactly the $K$ matrix for TQFT and the self-mirror symmetry condition \eqref{selfMirror} for SCFT. 
These conditions are indeed satisfied for self-mirror symmetry. Furthermore, the primitivity condition  for the embedding matrix is also satisfied because we  require that the charge matrix $Q$ should be primitive. Physically, this is the condition for the absence of 1-form symmetry in SCFT $\cT_Q$. 

So far, we have offered two general proofs for the relation between self-mirror symmetry and time-reversal invariance \eqref{selfduals}.  As we discussed before, we can also check the time-reversal invariance by computing the Gauss sum/generating function.  Although we are not able to prove that the Gauss sums/generating function are real under the condition $Q\Omega  Q^T=0$ in general, the statements can be verified in many examples, as we will show below. 


\subsection{Examples}
We now give some explicit examples to illustrate the relation between self-mirror symmetry  in SCFTs and time-reversal invariance in TQFTs.

Let us start with the simplest case of rank 1. The SCFT is obtained from the a single  U(1) gauge field coupled to matter and the TQFT is the $U(1)_k$ CS theory.
As we discussed in subsection~\ref{selfmirrorexample},  the self-mirror symmetric Abelian SCFT at rank 1 is given by $\cT_{Q=(m,n )}$,   classified by two coprime integers $m,n$. After performing the universal mass deformation, we get the CS theory $\mathscr T_{K=(m^2+n^2)}$, which is indeed time-reversal invariant  as spin TQFT after stacking with \sVec, as we discussed in subsection~\ref{CSTinvexample} and in particular in eq.~\eqref{tvalue}.
 
 The next example has rank 2. The self-mirror symmetric SCFT is specified by the   charge matrix  in  \eqref{rk2Q} and the resulting level matrix in TQFT can be easily computed    
 \be 
Q=\left(
\begin{array}{cccc}
 1 & 1 & 1 & 1 \\
 1 & 0 & 1 & 0 \\
\end{array}
\right)~, 
\qquad 
K=QQ^T=\left(
\begin{array}{cc}
 4 & 2 \\
 2 & 2 \\
\end{array}
\right)~.
\ee
 If we compute the Gauss sums $\tau_n(K)$ of $K$, we will find that they are not necessarily real. However, we can extend $K$ by adding 2 diagonal entries -1. Then the resulting extended matrix $\widehat K$   has only real Gauss sums, implying that $\sT_K$ is time-reversal invariant as  a spin TQFT. The Gauss generating function is given by 
 \be
 \Upsilon=\frac{1}{\sfz^2+1}+\frac{1}{1-\sfz}=2+\sfz+\sfz^3+2 \sfz^4+\sfz^5+\sfz^7+\cdots.
 \ee
 
 We finally consider an example with rank 3. The self-mirror symmetric SCFT is specified by the   charge matrix  in  \eqref{rk3Q} and the resulting level matrix in TQFT can be easily computed
\be
Q=\left(
\begin{array}{cccccc}
 1 & 0 & 0 & 1 & 0 & 0 \\
 0 & 1 & 0 & -1 & 1 & 0 \\
 0 & 0 & 1 & 0 & -1 & 1 \\
\end{array}
\right)~,
 \qquad
K=QQ^T=
\left(
\begin{array}{ccc}
 2 & -1 & 0 \\
 -1 & 3 & -1 \\
 0 & -1 & 3 \\
\end{array}
\right)~.
\ee
If we compute the Gauss sums $\tau_n(K)$ of $K$, we will find that they are not necessarily real. However, we can extend $K$ by adding 3 diagonal entries -1 (or equivalently one diagonal entry 1). Then the resulting extended matrix $\widehat K$   has only real Gauss sums, implying that $\sT_K$ is time-reversal invariant as  a spin TQFT. The Gauss generating function is given by 
\beqn
 \Upsilon=&=&\frac{\frac{2}{(-1)^{5/13} \sfz+1}+\frac{2}{(-1)^{7/13} \sfz+1}+\frac{2}{(-1)^{11/13} \sfz+1}+\frac{2}{1-(-1)^{2/13} \sfz}+\frac{2}{1-(-1)^{6/13} \sfz}+\frac{2}{1-(-1)^{8/13} \sfz}+\frac{1}{1-\sfz}}{\sqrt{13}}
\qquad
\\&=&\sqrt{13}+\sfz- \sfz^2+\sfz^3+\sfz^4-\sfz^5- \sfz^6-\sfz^7-\sfz^8+\sfz^9+\sfz^{10}-\sfz^{11}+\sfz^{12}+\cdots~.
\eeqn

\section{Conclusion} \label{conclusion}

In this paper, we have established an intriguing connection between two seemingly unrelated properties, mirror symmetry and time-reversal invariance. Focusing on the Abelian case, we show that self-mirror symmetric $\cN=4$ SCFTs yields time-reversal invariant TQFTs after performing the universal mass deformations. We discuss various methods to construct and discriminate self-mirror symmetric  SCFTs based on the charge matrices, Hilbert series and superconformal indices. We also review extensively various approaches to check the time-reversal invariance in TQFTs.  All these methods turn out to be useful and  enlightening to establish the connection between self-mirror symmetry and time-reversal invariance, which is proved from different perspectives for Abelian theories.

There are many interesting questions which remain  to be clarified and explored further. 
A very obvious and  interesting question is whether one can generalize the discussion in this paper to the non-Abelian case. One family of self-mirror symmetric non-Abelian SCFT is given by $T [SU(N)]$ \cite{Gaiotto:2008ak}, as we review in the main body. Meanwhile, it is worth mentioning that there is also a class of time-reversal invariant non-Abelian CS theory, given by $U(N)_{N, 2N}=(SU(N)_N\times U(1)_{2N^2})/\bZ_N$ where the quotient denotes the gauging of 1-form symmetry.  It would be interesting to understand whether  the $T[SU(N)]$   SCFTs and    $ U(N)_{N, 2N}$ TQFTs are related in some way.

It is also interesting to generalize the discussion in this paper to  those SCFTs  with 1-form symmetry, where the charge matrix is not primitive.   For this purpose, one has to establish systematically the mirror symmetry in this case which is more complicated due to the involvement of discrete gauge nodes. 

 There are also some conceptual puzzles regarding the connection between self-mirror symmetry and time-reversal symmetry. While the former is an internal symmetry, the latter is actually a spacetime anti-unitary symmetry.
 So the connection between them related by RG flow is novel and a bit surprising. It would be interesting to understand whether there is a deep reason or implication on the intrinsic structure of QFTs. Hopefully this would enable us to get a deeper understanding on those symmetries.

  We hope to study  these questions and report the progress in the near future. 
 
\acknowledgments

 We thank A. Hanany for discussions and D. Delmastro for correspondence. We also thank M. Buican, Z. Duan and A. Ferarri for previous related discussions.  
  This work was supported in part by the STFC Consolidated Grants ST/T000791/1 and ST/X000575/1.

\appendix

 \bibliographystyle{JHEP} 
 \bibliography {ref.bib}

  
\end{document}